\documentclass[apj,numberedappendix]{emulateapj}
\usepackage[english]{babel}
\usepackage{graphicx}
\usepackage{latexsym}
\usepackage{upgreek}
\usepackage{amsmath,amssymb}
\usepackage{natbib}
\usepackage{dcolumn}
\usepackage{bm}
\usepackage{calrsfs}
\usepackage{ulem}
\usepackage[usenames]{color}
\usepackage{xspace}

\begin{document}

\newcommand{\na}{New Astr.}
\newcommand{\nar}{New Astr. Rev.}
\newcommand{\jcap}{JCAP}

\def\be{\begin{equation}}
\def\ee{\end{equation}}
\def\ba{\begin{eqnarray}}
\def\ea{\end{eqnarray}}
\def\d{\delta}
\def\e{\epsilon}
\def\f{\varphi}
\def\k{\varkappa}
\def\tde{\tilde}
\def\p{\partial}
\def\ms{\mathstrut}
\def\s{\strut}
\def\ds{\displaystyle}
\def\ts{\textstyle}
\def\b{\boldsymbol}
\def\r{\mathrm}
\def\G{\Gamma}
\def\sun{\odot}

\def\ag{{\it AGILE}\xspace}
\def\fer{{\it Fermi} LAT\xspace}
\def\magic{MAGIC\xspace}
\def\hess{H.E.S.S.\xspace}
\def\veritas{{\it Veritas}\xspace}
\def\pks{\protect\object{PKS~2155$-$304}\xspace}
\def\opks{\protect\object{PKS~1510$-$089}\xspace}
\def\mkn501{\protect\object{Mkn~501}\xspace}
\def\ic{\protect\object{IC~310}\xspace}
\def\c454{\protect\object{3C~454.3}\xspace}
\def\cc{\protect\object{3C~279}\xspace}
\def\m87{\protect\object{M87}\xspace}

\title{Scenarios for ultrafast  gamma-ray variability in AGN}

\author{F.A.~Aharonian$^{1,2,3}$,
M.V.~Barkov$^{4,5,6}$, 
\and 
D.~Khangulyan$^{7}$
}

\affil{
$^{1}$Dublin Institute for Advanced Studies, 31
Fitzwilliam Place, Dublin 2, Ireland\\
$^{2}$Max-Planck-Institut f\"ur Kernphysik,
Saupfercheckweg 1, D-69117 Heidelberg, Germany\\
$^{3}$National Research Nuclear University “MEPhI”, Kashirskoje Shosse, 31, 115409 Moscow, Russia\\
$^{4}$Deutsches Elektronen-Synchrotron (DESY), Platanenallee 6, D-15738 Zeuthen, Germany\\
$^{5}$Astrophysical Big Bang Laboratory, RIKEN, 2-1 Hirosawa, Wako, Saitama 351-0198, Japan\\
$^{6}$Department of Physics and Astronomy, Purdue University, 525 Northwestern Avenue, West Lafayette, IN 47907-2036, USA\\
$^{7}$Department of Physics, Rikkyo University, Nishi-Ikebukuro 3-34-1, Toshima-ku, Tokyo 171-8501, Japan\\
}

\begin{abstract}
  We analyze three scenarios to address the challenge of ultrafast
  gamma-ray variability reported from active galactic nuclei. We focus
  on the energy requirements imposed by these scenarios: (i) external
  cloud in the jet, (ii) relativistic blob propagating through the jet
  material, and (iii) production of high-energy gamma rays in the
  magnetosphere gaps.  We show that while the first two scenarios are
  not constrained by the flare luminosity, there is a robust upper
  limit on the luminosity of flares generated in the black hole
  magnetosphere. This limit depends weakly on the mass of the central
  black hole and is determined by the accretion disk magnetization,
  viewing angle, and the pair multiplicity. For the most favorable
  values of these parameters, the luminosity for 5-minute flares is
  limited by \(2\times10^{43}\rm\,erg\,s^{-1}\),
  which excludes a black hole magnetosphere origin of the flare
  detected from \ic.  In the scopes of scenarios (i) and (ii), the jet
  power, which is required to explain the \ic flare, exceeds the jet
  power estimated based on the radio data. To resolve this discrepancy
  in the framework of the scenario (ii), it is sufficient to assume
  that the relativistic blobs are not distributed isotropically in the
  jet reference frame. A realization of scenario (i) demands that the
  jet power during the flare exceeds by a factor \(10^2\)
  the power of the radio jet relevant to a timescale of \(10^8\)
  years.
\end{abstract}


\maketitle

\section{Introduction}
\label{sec:intro} 
The hypothesis of supermassive black holes (SMBHs) as powerhouses of active galactic nuclei (AGN) has been proposed
\citep{sal64,zn66,lyn69} to explain the immense luminosities  of AGN and quasars by the release of the gravitational energy through the process
of gas accretion.  The radiation power of the accreting plasma is limited by the Eddington luminosity, 
\(L_{\rm Edd} = 1.3 \times 10^{46} M_8 \ \rm erg\,s^{-1}\), where 
\(M_8=M_{\rm bh}/10^8 M_\odot\) is the mass of the black hole in the unities of $10^8$ solar masses. 

The  apparent luminosities of radiation  of many AGN,  $L_{\rm app}=4 \pi D_L^2 f$ 
($f$ is the detected energy flux and $D_L$ is the source luminosity distance)  may exceed the  Eddington luminosity of SMBHs 
by  orders of  magnitude. However, the ``energy crisis''  can be overcome  if one assumes  
that the observed  radiation is highly anisotropic, namely,  that it is produced in  a collimated outflow  (jet) 
close to the line of sight \citep[see, e.g.,][]{bbr84}.

The concept of relativistically beamed emission offers not only  an elegant scheme for the unification of various classes of 
AGN \citep[see, e.g.,][]{up95},  but also provides a natural interpretation of  enormous fluxes of their nonthermal emission.  
Indeed, the assumption of production of radiation in a source relativistically moving toward the observer with a Doppler factor $\delta \gg 1$  allows physically reasonable  {\it intrinsic} luminosities of  AGN dubbed {\it blazers},  reducing
them by orders of magnitude compared to the apparent luminosity,   
 $L_{\rm app}=\delta^4 L_{\rm int}$.  This assumption also allows a larger (more ``comfortable'')  size of the production  
 region that is demanded by the observed variability of radiation: $l \leq c \Delta t_{\rm var} \delta$.   
These relations apply  to all  electromagnetic wavelengths, but they are crucial, first of all,  for gamma-ray loud  
AGN,  the apparent luminosities  of which during strong flares, e.g., in  3C454.3 \citep{agile10,fermi11_3C}  and  
\cc \citep{hnm15},   can achieve  the level of \(L_\gamma \sim 10^{49-50} \  \rm erg\,s^{-1}\).   Strong
Doppler boosting is also needed for prevent severe internal gamma-ray absorption, 
especially at VHE energies  \citep[see, e.g.,][]{cfr98}.

The gamma-ray emission of blazers is strongly variable, with fluxes that match the sensitivity of the 
{\it Fermi}  Large Area Telescope (LAT)  in the MeV/GeV band well, and the current arrays of imaging atmospheric Cherenkov telescopes 
(IACT), \hess, \magic, \veritas , in the VHE band.  During the strongest flares of BL Lac objects  like
Mkn~421, Mkn~501 and PKS~2155-304,  the energy fluxes  of VHE gamma-rays  often exceed 
\(f_{\rm VHE}=10^{-10} \  \rm erg\,cm^{-2}\,s^{-1}\).  Such fluxes can be  studied
with IACT arrays with huge detection areas that are as large as  $10^5 \ \rm m^2$ in almost background 
free regime,  allowing  variability studies on timescales of {\it minutes}.    
Although the  fluxes  of flaring  powerful quasars at MeV/GeV energies  can be significantly larger, 
\(f_{\rm VHE}=10^{-8} \  \rm erg\,cm^{-2}\,s^{-1}\),  because of the small  detection area 
of {space-borne instruments} ($\simeq 1 \ \rm m^2$), the capability  of the latter of probing the brevity of such 
strong AGN flares  has until recently  been limited by timescales of hours and days.  However,  
after the release of the  latest  software tools by the \fer collaboration,  
whivh allow a significant increase in the gamma-ray photon statistics, 
the variability studies at GeV energies for  exceptionally  bright flares  can be extended down to {\it minute} timescales.
This potential recently has recently been   demonstrated  by the \fer collaboration 
for  the  giant 2015 June outburst of 3C~273  \citep{aaa16}. 

It is straightforward to compare 
these timescales with the minimum  time that characterizes  a black hole system as an emitter, namely, the  light-crossing time of the gravitational radius of the black hole:
\be
\tau_0=r_{g}/c \approx 5\times10^2M_8\,\rm s. 
\ee
Note that $r_{g}=GM_{\rm bh}/c^2=1.5\times 10^{13}M_8\,\rm cm$  is 
the gravitational radius corresponding to the extreme
Kerr black hole, i.e.  twice smaller than the Schwarzschild radius. 

Thus, for the mass range of black holes $M \geq 10^8 M_\odot$, the current gamma-ray detectors 
have a potential to explore the physics of AGN that is close to the event horizon on timescales  shorter 
than $\tau_0$.  Such ultrafast  gamma-ray flares\footnote{For a recent summary of ultrafast gamma-ray flares of AGN   see \citet{vb15}}
have previously been detected from four AGN:
\pks \citep{ah07pks}, \mkn501 \citep{mkr501_magic},
and \ic \citep{ic310_14} at TeV energies, and \cc at GeV energies \citep{aaa16}. In addition, a flare with a duration comparable to the BH horizon 
light-crossing time, $\sim2\tau_0$, was observed from a missaligned radio galaxy \m87, in which the jet Doppler factor is expected to be 
small  \citep{gt09}. For comparison, it is interesting to note that the characteristic timescales of the even the shortest GRBs 
\citep[$\sim 1 \ \rm ms$;][]{pl02,gbl15}, which are most likely associated with solar mass black holes, exceed $\tau_0$ 
by several orders of magnitude.

The detection of variable VHE gamma-ray emission from AGN on
timescales significantly shorter than $\tau_0$ is an extraordinary
result and requires a careful treatment and interpretation. {The
  masses of SMBHs in distant AGN are typically derived from the
  empirical Faber-Jackson law \citep[also known as the \(M-\sigma\)
  relation, see][]{fm00,gbb00}. Although this statistical
  method is characterized by a small dispersion, scatter for individual objects may
  be significant, which consequently leads to uncertainties of \(\tau_0\).
  On the other hand, it follows from Eq.~\eqref{eq:t_var} that for the
  minute-scale flares reported from \pks and \ic, the variability time
  can exceed \(\tau_0\)
  only for masses of the BHs than are lower than
  \(3\times10^7\, M_\sun\).
  For both objects, different methods of estimating \(M_{\rm bh}\)
  give significantly higher values, and therefore \(\tau < \tau_0\). }

If the emission is produced in a relativistically
moving source with a velocity \(\beta_{\rm em}\), the variability
time-scale for the observer is shortened by the Doppler factor
\(\d_{\rm em}=1/\G_{\rm em}(1-\beta_{\rm em}\cos\theta_{\rm em})\);
\(\G_{\rm em}=1/\sqrt{(1-\beta_{\rm em}^2)}\) is the Lorentz factor and
\(\theta_{\rm em}\) is the angle between the source velocity and the
line of sight.  Thus if we wish to increase the proper size of the
emitter \(R'\) (the source size in the comoving reference frame) to a
physically reasonable value of \(R' \geq r_{g}\), the Doppler factor
should be large, \(\d_{\rm em}>10\). For example, in the case of
  \pks, where the mass of SMBH is expected to be high,\footnote{ To
    overtake this constraint, some models involve a BH binary system as
    the central engine in \pks \citep{rv10}} \(M_{8}\sim10\),
  the VHE variability sets a lower limit on the value of the Doppler
  factor: \(\d_{\rm em}\geq25\).  However, there is another issue of
conceptual importance that cannot be ignored.  The problem is that if
the perturbations originate in the central engine and then propagate
in the jet, e.g.  in the form of sequences of blobs ejected with
different Lorentz factors (leading to internal shocks), the size of
the emitter in the {\it laboratory} frame, \(R=R'/\G_{\rm j}\), would
not depend on the Doppler factor and it should exceed the
  gravitational radius: \(R\geq r_{g}\).  Let us present the proper
size of the production region as \(R'=\lambda\G_{\rm j}r_{g}\), where
\(\G_{\rm j}\) is the jet bulk Lorentz factor, and \(\lambda\) is a
dimensionless parameter, which corresponds to the ratio of the
production region size in the laboratory frame to the gravitational
radius.  The causality condition provides a limitation on the
variability timescale
\be\label{eq:t_var}
t_{\rm var}\geq\tau_0 {\lambda\G_{\rm j}\over \G_{\rm em}}\,.
\ee

The variability of $t_{\rm var}=0.04\tau_0$ inferred from the VHE flares of \pks \citep{ah07pks} requires
$\G_{\rm em}\simeq25\lambda\G_{\rm j}$, i.e., the emitter should move relativistically in the frame of the jet, which in
turn moves relativistically toward the observer.  The jet-in-jet model suggested by \citet{gub09} can be
considered as a possible realization of this general scenario. Alternatively, if the source of the flare does not move
relativistically relative to the jet ($\G_{\rm em} \simeq \G_{\rm j}$), the size of the source in the laboratory
  frame should be much smaller than the black hole gravitational radius: $\lambda\simeq0.04$.

If the emission site is located in the jet and formed by some perturbations propagating from the BH, one should expect
$\lambda>1$. Thus, the condition of $\lambda<1$ implies that the perturbations in the jet that result in a flare should
have an {\it external} origin, i.e., are not directly linked to the central black hole. This scenario can be realized when a
star or a gas cloud of radius $R_*\ll r_{g}$ enters the jet from outside and initiates perturbations on scales
smaller than the black hole gravitation radius $r_{g}$ \citep{babkk10}.

Finally, it has been suggested that the flares can be produced in the BH magnetosphere \citep{NerAha07,lr11,rieger11}. In this
case, the production site does not move relativistically with respect to the observer, and Eq.~\eqref{eq:t_var} is reduced
to $t_{\rm var}>\tau_0\lambda$, where $\lambda=R/r_{g}$. Thus, the flare originates in a compact region that
occupies a small fraction of the black hole magnetosphere. An analogy for this possibility could be the emission of
radio-loud pulsars.  It is believed that in these objects the radio pulses are produced in the polar cap region, which
constitutes only a small part of the pulsar surface. Note that for the typical pulsar radius $R_{\rm psr}$ of 10 km,
$\tau_0=R_{\rm psr}/c \sim 30\rm \upmu s$ is too small to be probed through the variability of the radio emission. We note
here that although the production site of relativistic motion does not allow to reducing the minimum variability time (see
Eq.~\eqref{eq:t_var}), the relativistic beaming effect allows significant relaxation of the energetics required to
produce the flare. Thus, magnetospheric scenarios should have higher energy requirements then the jet scenarios.

In this paper we discuss in rather general terms  
three possible scenarios for the production of ultrafast ("subhorizon" scale)  variability in AGNs: 
\begin{itemize}
\item[(i)] The source of the flare is a magnetospheric gap  occupying a small volume in the proximity 
of the black hole close to the event horizon \citep{NerAha07,lr11}.
\item[(ii)] The emitter  moves relativistically in the jet reference frame. The most feasible energy
source for this motion is magnetic field reconnection in a  highly magnetized jet \citep{lub05,gub09,pgs16}. 
\item[(iii)] Flares are  initiated by  penetration  of   external  objects (stars or clouds)  
into the jet \citep{abr10,babkk10}.   
\end{itemize}

Apparently,  any model designed to explain the ultrafast variability  on timescales 
$t_{\rm var}<\tau_0$  should address some other key issues. In particular,  
the  required overall energy budget should be feasible,   the  source 
should be optically thin for  gamma-rays, and of course,  
the proposed radiation mechanism(s)  should  be able to explain the 
reported spectral  features of gamma-ray emission.

\section{Addressing the  ``subhorizon'' scale variability}
\label{sc:energy}

\subsection{The Magnetospheric Model}

Magnetospheres of the central SMBHs in AGN can be sites of production of gamma-rays with spectra extending to VHE energies 
\citep[see, e.g.,][]{BIP92,lev00,NerAha07,RieAha08,lr11}. At low accretion rates, the injection of charges into the BH magnetosphere 
is not sufficient for a full screening of the electric field induced by the rotation of the compact object. The regions with unscreened 
electric field, referred to as gaps, are capable of effective acceleration of charged particles. Such a scenario may result in a variability 
of the source on ``subhorizon'' timescales since the size of the gap is much smaller than the gravitational radius. The attractiveness of 
this scenario is its applicability to the non-blazar-type AGN.  On the other hand, because of both the low accretion rate and the lack of 
Doppler boosting,  the gamma-ray luminosities of such objects are expected to be quite modest when compared to blazers. Therefore  
the detectability of the black hole magnetospheric radiation is most likely limited by a few nearby objects. In particular, the radio galaxy 
M87, as well as the compact radio source Sgr A* in the center of our Galaxy, can be considered as suitable candidates for the realization 
of such a scenario \citep[see, e.g.,][]{lr11}.

The energy release in the entire magnetosphere is limited by the BZ
luminosity. Below we follow a simplified treatment that allows us to
estimate the energy release in a thin vacuum gap formed in the SMBH
magnetosphere.  The rotation of a magnetized neutron star or BH in
vacuum induces an electric field, $\mathbf{E_0}$, in the surrounding
space \citep{gj69,BZ77}.  If a charge enters this region, the
electric field should accelerate it.  In an astrophysical context the
unscreened electric field is usually strong enough to boost the
particle energy to the domain where the particle starts to interact
with the background field and thus initiates an electron-positron
pair cascade. The secondary particles move in the magnetosphere in a
way that tends to screen the electric field
\citep{stu71,RudSut75}. Eventually, an electric-field-free
configuration of the magnetosphere can be formed.  However, one
  should note that there are differences between the structures of the
  pulsar and BH magnetospheres, and consequently, the theoretical
  results obtained for pulsar magnetospheres cannot be directly
  applied to the BH magnetosphere. In particular, while in the case of
  the pulsar magnetosphere the source of the magnetic field is well
  defined, in the BH magnetosphere the magnetic field is generated by
  currents in the disk and magnetosphere. The configuration of the
  field is determined by the structure of the accretion flow. Thus, a
  change of the accretion flow can result in the formation of
  charge-starved regions (gaps) in the BH magnetosphere. 

The charge density required for the screening is known as the
Goldreich-Julian density (GJ), $\rho_{\rm GJ}$ \citep{gj69}.  However, the
process of the pair creation is expected to be highly non-stationary
\citep[see, e.g.,][ for theoretic and numerical considerations of the pair creation in
vacuum gaps, respectively]{lmj05,timokhin10}, thus  even if am electric-field-free 
state of the magnetosphere is possible, it cannot be
  stable \citep{stu71}. The gaps, i.e., regions in which the charge density is not
sufficient for the electric field screening, may appear sporadically
in the magnetosphere, for example, in the vicinity of the
  stagnation surface, i.e., at the boundary that separates acretion and ejection
  trends in the flow.

The vacuum electric field strength $\mathbf{E_0}$ determines the maximum electric
field in the gap, thus the maximum acceleration rate of a particle with charge $e$ in the gap is
$mc^2\dot{\gamma}< e c E_0$. The total power of particle acceleration can be expressed as
\be\label{eq:e_dot}
\dot{\cal E}<\int\limits_{\rm gap}dV\,e (n_{e}+n_{e^{+}}) c E_0\,,
\ee
where $n_{\rm e}$ and $n_{e^{+}}$ are densities of electrons and positrons. If all the energy gained by the particles in the gap is emitted in gamma rays, 
Eq.~\eqref{eq:e_dot} also corresponds to the upper limit of the gamma-ray luminosity. Note that electrons and positrons move in opposite directions  in the gap, 
and only one of these species generates emission detectable by a distant observer.   

For a thin spherical gap, $R<r<R+h$, the luminosity upper limit is
\be 
L_\gamma<4\pi R^2 h e n_{\rm e} c E_0\,,
\ee
where the electrons are assumed to emit outward.  The particle density can be expressed as a fraction of the
Goldreich-Julian density: $e n_{\rm e}=\kappa \rho_{\rm GJ}$, where $\kappa$ is the multiplicity. The condition for the
electric field screening, \(e|n_{\rm e}-n_{\rm e^+}|=\rho_{\rm GJ}\), allows charge configurations with high multiplicity
and still non-screened electric field. To obtain a more detailed estimate of the generated pairs' influence on the
electric field in the gap, it is necessary to consider the electromagnetic cascade in the gap.

The numerical simulations of \cite{timokhin10} and \cite{ta13} demonstrate an important tendency. When the multiplicity becomes significant,
$\kappa\sim1$, the charges in the gap start to generate an electric field that is comparable to $E_0$, and the accelerating field
vanishes.  Thus, for effective charge acceleration, the following condition should be fulfilled: $\kappa \ll 1$.  Thus, the
total energy release in the gap of thickness, $h$, can be estimated as
\be\label{eq:gamma_lum}
L_\gamma<4\pi R^2 h\kappa \rho_{\rm GJ} c E_0\,.  
\ee
The electrical field in the gap is estimated as
\be \label{eq:E_0}
E_0 \approx  B_{\rm g}{R\Omega_{\rm F} \sin\theta\over c}\,, 
\ee
where $B_{\rm g}$ is the magnetic field in the vacuum gap, $\Omega_{\rm F}$ is the angular velocity of the frame, $R$ is
the radius, and $\theta$ is the polar angle. {In fact, the actual electric field in the drop is smaller by a factor
$h/R$ than the value given by Eq.~\eqref{eq:E_0} \citep[see, e.g.][]{BZ77,lr11}. This factor accounts for the influence 
of the magnetospheric charges located outside the gap.  Eq.~\eqref{eq:E_0} does not account for this contribution. Since these charges, even 
if they remain outside the gap, tend to decrease the electrical field in the gap, Eq.~\eqref{eq:E_0} provides a strict upper limit on the gap electric field strength.

The Goldreich-Julian density is also determined by the same parameters:
\be
\rho_{\rm GJ}=\Omega_{\rm F}B_{\rm g}\sin\theta/(2\pi c)\,.
\ee
For a Kerr BH  with the maximum angular momentum,  the angular velocity $\Omega_{\rm F}$ is 
estimated as
\be
{\Omega_{\rm F}\over c}\simeq \frac1{4r_g}\,.
\label{eq:omega}
\ee
Substituting  Eq.~(\ref{eq:E_0})~-~(\ref{eq:omega}) to Eq.~\eqref{eq:gamma_lum}, one obtains
\be
L_\gamma<\frac18{{B_{\rm g}^2 R^3\over r^2_{\rm g}}\kappa h c\sin^2\theta}\,. 
\label{lgmax_o}
\ee
{The upper limit on the luminosity from a vacuum gap depends on the factor $R^3B_{\rm g}^2$ that is expected to decrease with $R$.  
For sake of simplicity {below it is adopted } that  $R^3B_{\rm g}^2\simeq r_{g}^3B_{\rm bh}^2$, where $B_{\rm bh}$ is the magnetic field at the BH horizon. Thus, one obtains}
\be
L_\gamma<\frac18{ B_{\rm bh}^2 r_{g} \kappa h c\sin^2\theta}\,. 
\label{lgmax}
\ee
We should note that for \(h \rightarrow r_{g}\), the luminosity estimate provided by Eq.~(\ref{lgmax}) (after
averaging over the polar angle \(\theta\)) exceeds the Blandford-Znajek (BZ) luminosity
\citep{BZ77,beskin_book} by a factor of $2$. This is imposed by several simplifications in our treatment.
The most important contribution is caused by the
usage of the electric field upper limit, Eq.~\eqref{eq:E_0}, as the accelerating field. 

Thus, Eq.~(\ref{lgmax}) can be considered as a safe upper limit for the luminosity of magnetospheric flares.  A similar
estimate has been obtained by \cite{rieger11} and \cite{lr11}. However, the numerical expression in \citet{lr11} 
contains some uncertain geometrical factor \citep[\(\eta\) in the notations of][]{lr11}. Eq.~(\ref{lgmax}) allows us to 
estimate its value: this geometrical factor should be small, \(\sim10^{-2}\) \citep[see also Eq.~52 in][]{rieger11}. 

Finally, \citet{BroTch15} {argued} that for the full screening of the electric field in a thin gap, the charge density should exceed 
the Goldreich-Julian value by a factor\footnote{Note that \citet{BroTch15} used a different notation for the gap thickness, \(\Delta\).} 
\(R/h\), which should lead to an enhancement of the gap radiation. To illustrate the physical reason for the existence of this factor, 
\citet{BroTch15} computed the divergence of the electric field in the gap. However, as the gap electric field they adopted a field determined 
by an expression similar to Eq.~\eqref{eq:E_0}, i.e., a value that overestimates the true field by the factor \(R/h\) 
\citep[see, e.g.,][for a more accurate introduction of the electric field in the gap]{BZ77,lr11,rieger11}. 
Thus, the factor suggested by \citet{BroTch15} seems to be strongly overestimated.

Finally, the thickness of the gap, $h$, in Eq.~\eqref{lgmax} is constrained by the  variability time scale, 
$h \sim t_{\rm var}c$.
To production the emission variable on a 5-minute time-scale, $t_{\rm var}= 5\,t_{\rm var,5}\rm\,min$, 
the gap thickness, $h=10^{13}t_{\rm var,5}\rm \,cm$, should be smaller than the gravitational radius of the SMBH with a mass $M_8>1$. Thus,
the estimated gamma-ray luminosity cannot exceed the following value:
\be
\label{eq:L_max}
L_\gamma<5\times10^{43}\kappa B_{4}^2 M_{8}t_{\rm var,5}\sin^2\theta \rm\,erg\,s^{-1}\,,
\ee
where $B_{\rm bh}=10^4B_4\,\rm G$.

Eq.~\eqref{eq:L_max} contains two parameters that are determined by properties of the advection flow in the close vicinity of the BH: 
pair multiplicity, $\kappa$, and the magnetic field strength, $B$. Importantly, these parameters are essentially defined
by the same property of the flow, more specifically, by the accretion rate. The magnetic field at the BH horizon needs to be supported by 
the accretion flow. Therefore the field strength is directly determined by the accretion rate. The accretion rate also defines 
the intensity of photon fields in the magnetosphere, and consequently, the density of electron-positron pairs produced through gamma-gamma 
interaction \citep[see, e.g.,][]{lr11}. If the multiplicity parameter, $\kappa$, approaches unity, the gap electric field vanishes 
\citep[see, e.g.,][]{ta13}.  This sets an upper limit on the accretion rate, and consequently on the magnetic field strength.

In previous studies \citep{lr11,ic310_14} the maximum accretion rate compatible with the existence of a vacuum gap in the magnetosphere 
was estimated as
\begin{equation}\label{mdot_max}
\dot{m}<3\times10^{-4}M_{8}^{-1/7}\,,
\end{equation}
where $\dot{m}$ is the accretion rate in the Eddington units:
\be
\dot{M}_{\rm edd} = \frac{4\pi m_p G M_{\rm bh}}{\eta c \sigma_t}\,.
\label{mdot_edd}
\ee
Here \(m_p\), \(\sigma_t\), and \(\eta\) are the proton mass, the Thompson cross-section, and the accretion efficiency factor, respectively. 

To derive the estimate provided by Eq.~\eqref{mdot_max}, \citet{lr11} adopted a value of $\eta=0.1$
and estimated the magnetic field strength at the BH horizon as
\begin{equation}
B_{\rm bh}=1.3\times10^5\left(\dot{m}/M_8\right)^{1/2}\rm\, G\,,
\end{equation}
where we rescaled the numerical coefficient to the normalization used throughout our paper. 
For this magnetic field strength, Eq.~\eqref{eq:L_max} yields
\be
L_\gamma<3\times10^{42}\kappa M_{8}^{-1/7}t_{\rm var,5}\sin^2\theta \rm\,erg\,s^{-1}\,.
\ee
 In some cases, e.g., for \ic, the energy requirements are rather close to the obtained upper limit, therefore we consider a somewhat
  more accurate treatment of the case of a magnetosphere around a Kerr BH below.

The strength of the magnetic field at the BH horizon can be obtained by extrapolating the field at the last marginally stable orbit. 
Let us define the magnetic field in the disk as
\be\label{eq:b_disk}
B_{\rm d}=\sqrt{8\pi\beta_{\rm m} p_{\rm g}}\,,
\ee
where $\beta_{\rm m}$ and $p_{\rm g}$ are the disk magnetization and gas pressure in the accretion disk that confines the magnetic
field at the horizon.  The gas pressure can be estimated using the solution for a radiatively inefficient accretion flow
\citep[see ][]{ny94} \footnote{A more accurate treatment of the accretion flow {reveals} a correction by less than 30\%  \citep[see
][]{ny95a} as compared to the height-averaged treatment in \citet{ny94}} as
\be
p_{\rm g}=\frac{\sqrt{10}\dot{M}\sqrt{G M_{\rm bh}}}{12\pi \alpha_{ss} R^{5/2}}\,,
\ee
where $\alpha_{ss}$ is the nondimensional viscosity of the disk \citep{ss73}.
{Eq.~\eqref{eq:b_disk} for $R\rightarrow r_{g}$ provides an estimate for the magnetic field at the BH horizon}:
\be
B_{\rm bh}=1.5\frac{ \beta_{\rm m}^{1/2} (\dot{M}c)^{1/2}} {(\alpha_{ss})^{1/2} r_{g}}\,.
\label{bpp}
\ee

The magnetic field strength provided by Eq.~\eqref{bpp} together with Eq.~\eqref{lgmax} yields in
\be
L_{\gamma} < \frac{\sqrt{10}}{12}\frac{\beta_{\rm m}\kappa (h/r_g)\sin^2\theta \dot{M} c^2}{\alpha_{ss} }\,.
\label{lgam_m}
\ee

The multiplicity parameter, $\kappa$, at the Kerr radius is determined as ({ see Appendix \ref{app:npm} for details})
\be
\kappa \equiv\frac{n_{\pm}}{n_{\rm GJ}} \approx 6\times10^6\frac{ \dot{m}^{7/2}M_{8}^{1/2}}{(\eta \alpha_{ss})^{7/2}\beta_{\rm m}^{1/2}}.
\label{kappa_kerr}
\ee
The condition $\kappa<1$ determines an upper limit on the accretion rate:
\be
\dot{m}<10^{-2}\frac{\eta \alpha_{ss} \beta_{\rm m}^{1/7}}{M_{\rm bh}^{1/7}}.
\label{mdot_max_kerr}
\ee
Eq.~(\ref{mdot_max_kerr}) and (\ref{mdot_edd}) substituted into  Eqs.~(\ref{bpp}) and (\ref{lgam_m}) give an upper limit for the magnetic field that is consistent with the existence of vacuum gaps:
\be
B_{\rm bh}< 7\times 10^3 \left(\frac{\beta_{\rm m}}{M_{8}}\right)^{4/7}\mbox{ G}\,,
\label{bpp_max}
\ee
and consequently, the maximum luminosity of particles accelerated in the gap does not depend on $\alpha_{ss}$ and $\eta$:
\be
L_{\gamma} <  2\times10^{43} \beta_{\rm m}^{8/7} \kappa t_{\rm var,5} M_{8}^{-1/7} \sin^2\theta \,\mbox{ erg s}^{-1}\,.
\label{lgmm}
\ee
This estimate is obtained for the thick-disk accretion
  (in the ADAF-like regime). The limit on the accretion rate given by
  Eq.~\eqref{mdot_max} is consistent with the realization of this
  accretion regime. For higher accretion rates, \(\dot{m} \geq 0.1\),
  the accretion flow is expected to converge to the thin-disk solution
  \citep{BisBli77,acl88}. In this regime, the temperature of the disk
  is expected to be significantly below \(1\rm\, MeV\),
  thus the pair creation by photons supplied by the accretion disk
  should be cease. This effectively mitigates the constraints
  imposed by the accretion rate. However, the change of the accretion
  regime also significantly weakens the strength of the magnetic field
  at the BH horizon \citep{BisLov07}, and consequently decreases the
  available power for acceleration in the gap.

To derive  Eq.\eqref{lgmm}, we assumed that the gap thickness is determined by the variability time-scale; this corresponds to the energetically 
most feasible configuration. In a more realistic treatment, one should also take into account the interaction of the particles that are accelerated 
in the gap with the background radiation field. For high and ultrahigh energies of electrons, \(E>1\rm \,TeV\),   the characteristic time of 
the inverse Compton scattering appears to be shorter than the minute-scale typical for the short TeV flares (see Appendix~\ref{app:npm}).   
For the hot target photon field, as expected from a thick accretion disk, the pair-production process should also be very efficient, 
\(\lambda_{\gamma\gamma}\leq\lambda_{\rm IC}\). Thus, computation of the TeV emission requires a detailed modeling of the electromagnetic 
cascade \citep[see, e.g.,][]{BroTch15,hpc16}. Furthermore, the production and evacuation of the cascade-generated pairs may follow a cyclic 
pattern and the inductive electric field may become comparable to the vacuum field \citep{lmj05}. A detailed consideration of this complex 
dynamics is beyond the scope of this paper, but we note that the characteristic length of such a cascade-moderated gap should be small, 
\(\sim\sqrt{\lambda_{\rm IC}\lambda_{\gamma\gamma}}\), resulting in a reduction of the available power \citep[see also][]{BIP92}.

Eq.~\eqref{lgmm} determines the maximum luminosity of vacuum gaps that can collapse quicker than $t_{\rm var}$. It
has been assumed for its derivation that the magnetic field is determined by an accretion regime that its in turn determines the
intensity of the photon field in the magnetosphere. In the case of a steady accretion, this seems to be a very feasible approximation. However, 
this may look less certain in the case of a rapidly changing accretion rate, since the processes that govern the variation of the accretion rate 
and escape of the magnetic field {from the BH horizon} may have different characteristic timescales.  Therefore we provide some estimates 
for these two timescales below. 

The dominant contribution to the photon field comes from plasma located at distances $r\sim 2r_g$, and the
characteristic viscous accretion  time (density decay time in the flow) is
\be
t_{\rm \rho, decay}\simeq {2r_{g}\over c\alpha_{\rm ss}}\simeq10^4\alpha_{\rm ss,-1}^{-1}M_{8}\rm \,s\,.
\ee
When the accretion fades, the decay of the magnetic field is determined by the magnetic field reconnection
rate \citep{kom04}:
\be
t_{\rm B, decay}\simeq {\pi r_{g}\over 0.3 c}\sim10^4M_{8}\rm \,s\,.
\ee
Since these two time-scales are essentially identical, it is natural to expect that the field strength and the disk density
will decay simultaneously. Thus, Eq.~\eqref{lgmm} should also be valid for {the time-dependent accretion regime}.

\subsection{Relativistically Moving Blobs}
The properties of radiation generated in jets may be significantly affected if some jet material moves relativistically
with respect to the jet local comoving frame. For example, the magnetic field reconnection may be accompanied by
the formation of slow shocks \citep[see, e.g.,][]{lub05} that in the magnetically dominated plasma produce relativistic
flows \citep{kom03}. If such a process is realized in AGN jets, it can lead to gamma-ray flares in blazar-type AGN with
variability timescale significantly shorter than $r_{g}/c$ \citep{gub09}.  Another implication of this scenario is
related to short gamma-ray flares detected from missaligned radio galaxies \citep{gub10}. Indeed, the conservation of
momentum requires that for each plasmoid directed within the jet-opening cone, there should exist a counterpart
that is directed outside the jet-beaming cone. { While the radiation of the plasmoid directed along the jet appears as a short
  flare, the emission associated with its counterpart outflow can be detected as a bright flare by an off-axis
  observer. The latter process may have a direct implication on the interpretation of flares from nearby missaligned
  radio galaxies, e.g., \m87 \citep{gub10}}.

If a process, operating in a region of the jet with comoving volume $V'$, results in the ejection of plasmoids, 
some fraction, $\xi$, of the energy contained in the volume is transferred to the outflow. The conservation 
of energy can be written as
\be
\xi V' \epsilon'_{\rm j}= S_{\rm co}\G_{\rm co}^2 v_{\rm co}\Delta t' (4/3\tilde{\epsilon}_{\rm e})\,.
\ee
Here $\epsilon'_{\rm j}$ and $\Delta t'$ are the energy density of the jet plasma and duration of the ejection, as seen in the jet 
comoving reference frame. $\G_{\rm co}=1/\sqrt{1-(v_{\rm co}/c)^2}$, $S_{\rm co}$, and $\tilde{\epsilon}$ are the plasmoid Lorentz factor, 
the outflow cross-section, and the internal energy, respectively. The outflow  cross-section can be estimated as 
$S_{\rm co}\simeq S/(2\G_{\rm co}^2)$, where $S\simeq V'/(\Delta t' v_{\rm co})$ is the surface of the volume $V'$. Thus, 
one obtains an estimate for the energy density of the 
\be
\epsilon'_{\rm j}\simeq\frac2{3\xi}\tilde{\epsilon}_{\rm e}\,.
\ee
For simplicity, in what follows we take $\xi\epsilon'_{\rm j}\simeq\tilde{\epsilon}_{\rm e}$. The efficiency of the energy transfer, 
$\xi$, depends on a specific realization of the scenario. 
For example, it seems that the for the  reconnection of the magnetic field, the efficiency might be high $\xi\sim1$, 
as follows from  an analytic treatment by \citet{lub05} and 
  the results of numerical simulations by \citet{sgp16}\footnote{From Figure 2 of \citet{sgp16} it follows that
    $n_{\rm lab}\simeq\sigma n_0$ and $<\gamma>_{\rm lab}n_{\rm lab}\simeq\sigma^2 n_0$, thus the internal energy in
    the plasmoid is  $\tilde\epsilon\simeq \tilde{n}<\tilde{\gamma}>\simeq\sigma n_0mc^2$.}. 
    We note, however, that in the presence of a guiding field, 
    the magnetization of the ejected plasmoids should be high \citep{lub05}.
  
On the other hand, the energy density in the plasmoid can be estimated through the
emission variability time and the luminosity level \citep[see Eq.~(9)][]{gub09}:
\be
 \tilde{\epsilon}_{\rm e}={E_{\rm em}\over \G_{\rm em}\tilde{l}_{\rm em}^3}\,,
\ee
where the variability time-scale determines the size of the production region: $\tilde{l}_{\rm em}=c\Delta t \G_{\rm em}$, and
the flux level defines the energy content in the plasmoid: $E_{\rm em}=L_{\gamma}\Delta t/(4f\G_{\rm em}^2)$ (here $f<1$
defines the fraction of the plasmoid energy transferred to the flare emission; this factor is dropped in what
follows, {and its contribution is accounted for in the value of the factor $\xi$}). Thus, one obtains
\be
 \tilde{\epsilon}_{\rm e}={L_{\gamma} \over 4\G_{\rm em}^6c^3\Delta t^2}\,.
\ee
On the other hand, the energy density in the jet is
\be
 {\epsilon'}_{\rm j}={L_{\rm j} \over \Delta\Omega r^2 c \G_{\rm j}^2}\,,
\ee
where $\Delta\Omega\simeq \pi/\G_{\rm j}^2$ is the jet propagation solid angle. 
The comparison of these equations allows us to estimate the required true luminosity of the jet as 
\be
L_{\rm j}={L_{\gamma}\over \G_{\rm em}^{6}}{\pi r^2\over 4\xi c^2\Delta t^2}\,.
\ee
{The above equation is consistent with Eq.~(10) from \citet{gub09}. Note, however, a difference in the notations:
throughout this paper, $L_{\rm j}$ is the true jet luminosity, while in \citet{gub09} $L_{\rm j}$ corresponds to the
  isotropic luminosity.}

If the viewing angle is small, the mini-jet Lorentz factor can be expressed as 
$\G_{\rm em}=2\G_{\rm j}\G_{\rm co}/(1+\alpha^2)$ where $\alpha=\theta\G_{\rm j}$ is the viewing angle 
expressed through the jet-opening angle (see Appendix \ref{app:sal}). 
Thus, the above equation can be simplified as
\be
L_{\rm j}={L_{\gamma}\over \G_{\rm co}^{6}\G_{\rm j}^6}{\left(1+\alpha^2\right)^{6}\over256}{\pi r^2\over \xi c^2\Delta t^2}
\ee
or
\be\label{eq:j_in_j_lum}
L_{\rm j}=1.4\times10^{-5}L_{\gamma}\left({1+\alpha^2\over 4}\right)^{6}\G_{\rm co,1}^{-6}\G_{\rm j,1}^{-6}\xi_{-1}^{-1}r_2^2M_8^2t_{\rm var,5}^{-2}\,.
\ee
Here it was assumed that the flare originates at a distance $r_2=100 r_{g}$ from the central BH with mass $M_{\rm bh}=10^8M_\odot M_8$. 

The above estimate describes the jet luminosity requirement to generate a single short flare of duration \(t_{\rm var}\).
Observations in HE and VHE regimes show that AGNs often demonstrate a rather long period of activity (as compared to the 
duration of a single peak): \(T\gg t_{\rm var}\). If the mini-jets are isotropically distributed in the jet comoving frame,
the probability for an observer to be in the mini-jet beaming cone depends weakly on the observer viewing 
angle\footnote{If $\Gamma_{\rm co}>\Gamma_{\rm j}$, this statement is correct for observers located in 
$\theta_{\rm view}<\pi/2$, otherwise for $\tan\theta_{\rm view}<v_{\rm co}/\sqrt{1/\G_{\rm co}^2-1/\G_{\rm j}^2}$.}, 
and this probability can be estimated as 
$P\simeq(2\G_{\rm co})^{-2}$ \citep{gub10}. If the mini-jet formation is 
triggered by some {\it spontaneous} process, then the
comoving size of the region responsible for the flare is $l'_0=\d_{\rm j}Tc$, and the energy contained in this region is
$E'=Sl'_0e'_{\rm j}$ (here $S$ is the jet cross-section). The energy of a single mini-jet in the comoving frame is
\be
E'_{\rm mj}={L_{\gamma}t_{\rm var}\G_{\rm co}\over 4\xi\G_{\rm em}^3}\,.
\ee
The total number of mini-jets during a flaring episode can be estimated as $N\approx \Phi T/P t_{\rm var}$, 
where $\Phi$ is the so-called filling factor.

The total dissipated energy for the flare should be smaller than the energy that is contained in the dissipation region:
\be
{E'_{\rm mj}\Phi T\over Pt_{\rm var}} <  L_{\rm j}T{\d_{\rm j}\over \G_{\rm j}^2}\,.
\ee
This implies a requirement for the jet luminosity 
\be\label{eq:j_in_j_lum_total}
L_{\rm j}> 0.1\Phi \zeta^2 \d_{\rm j,1}^{-2} L_{\gamma} \xi_{-1}^{-1}\,,
\ee
here $\zeta=\G_{\rm j}/\d_{\rm j}$, or 
\be\label{eq:j_in_j_lum_total2}
L_{\rm j}> 0.006 \Phi\left({1+\alpha^2}\right)^4\G_{\rm j,1}^{-2}L_{\gamma} \xi_{-1}^{-1}\,,
\ee
where the small viewing angle limit was used for the ratio of Lorentz
and beaming factors:
$\zeta=\G_{\rm j}/\d_{\rm j}\simeq(1+\alpha^2)/2$ (see appendix~\ref{app:sal}). The requirement
imposed by Eq.~\eqref{eq:j_in_j_lum_total} significantly exceeds the
limit related to the shortest variability time, Eq.~\eqref{eq:j_in_j_lum}. 

{The derived lower limit for the jet luminosity contains the parameter $\xi$, which accounts for the conversion efficiency
  from jet material to the outflow, and from outflow to the radiation. While the letter can be high, $\sim1$, if a good
  target for nonthermal particles exists, the value of the former efficiency depends on the process behind the outflow
  formation. For example, it was argued that if the outflow is formed by the Petschek-type relativistic reconnection \citep{lub05}, the energy 
transfer is expected\footnote{This follows from Eq.(6) in \citep{gub09}, i.e., $\tilde{\epsilon}_{\rm e}= \sigma \rho_{\rm j}'c^2\simeq \epsilon_{\rm j}'$ for $\sigma\gg1$.} to be high, $\sim1$. 
  However, the efficiency of the transfer can be significantly suppressed if the guiding field is present in the reconnection domain \citep{lub05,bk16}.  }

On the other hand, this requirement can be somewhat relaxed if the velocity direction of the plasmoids is not
random, e.g., is controlled by the large-scale magnetic field \citep{gub09}, or is triggered by some perturbation
propagating from the base of the jet. In the former case the mini-jet detection probability, $P$, may be higher, and in
the latter case, the comoving distance between the mini-jets may be larger than $l'_0$. Let us assume that the flare
trigger propagates with Lorentz factor $\G'_{\rm tr}$ in the jet comoving frame, then the comoving region size is 
larger by a factor of $\G'_{\rm tr}$.

\subsection{Cloud-in-Jet Model}

In the framework of the cloud-in-jet scenario, we deal with the nonthermal emission generated at the interaction of a
jet with some external obstacle, e.g., a BLR cloud or a star \citep[see, e.g.,][]{bk79,bp97,abr10,bab10,bba12}. Debris of
the obstacle matter, produced at such an interaction, can be caught by the jet flow. This debris should form dense
blobs or clouds in the jet, and the emission generated during their acceleration may be detected as a flare
\citep{babkk10,kbbad13}.  If this interpretation is correct, each peak of the light curve can be associated with emission
produced at the acceleration of some individual blob.  The peak profile and its duration are determined by the condition of how
quickly this blob can be involved into the the jet motion, i.e., by the dimension and mass of the blob. Light blobs with a
 mass satisfying the condition
\be\label{eq:light_blob}
M_{\rm c} c^2 <  {P_{\rm j}\pi R_{\rm c}^2r_0 \over 4\G_{\rm j}^3}\,
\ee
(here \(P_{\rm j}\) and \(R_{\rm c}\) are the jet ram pressure and the cloud radius, respectively)  are accelerated on length scales smaller than the distance to the SMBH, \(r_0\) \citep[for a consideration of heavy blobs that can be accelerated over a distance comparable to \(r_0\), see][]{kbbad13}.

If this condition is fulfilled, the variability time scale can be estimated as \citep{babkk10,kbbad13}
\be \label{eq:t_var_sij}
t_{\rm var}\simeq {4cM_{\rm c} \G_{\rm j}^2\over P_{\rm j}\pi R_{\rm c}^2 \d_{\rm j}}\,.
\ee
This estimate ignores perturbations in the jet that are generated by the obstacle-jet interaction, which is probably very complex,
and their influence can be explored only by numerical simulations, which are beyond the scope of this paper \citep[see, e.g.,][]{cbp16}. 

In the case of a cloud with a mass that satisfies to Eq.~\eqref{eq:light_blob}, the variability can be very short:
$t_{\rm var}<r_0/(c\G_{\rm j}\d_{\rm j})$, but obviously it cannot be shorter than the light-crossing time of the blob:
$t_{\rm var}>R'_{\rm c}/(c\d_{\rm j})$. This implies, through Eq.~\eqref{eq:t_var_sij}, that the minimum mass of blobs below which the dynamics of the blobs
is very quick and the variability is limited by the blob light-crossing limit.

On the other hand, the mass of the blob determines the energy transferred by the jet to the blob during its
acceleration, and consequently, the apparent energy emitted in the corresponding peak of the light curve
\citep{kbbad13}:
\be\label{eq:energy_sij}
E_{\gamma}\simeq {\xi M_{\rm c}c^2 \d_{\rm j}^3 }\,,
\ee
where the factor $\xi$ accounts for the fraction of energy transferred to the gamma-ray emitting particles and some dynamical
factor \citep[$\sim0.3$, which is related to the {\it correction function} $F_{\rm e}$ defined in ][]{babkk10,kbbad13}.
Thus, if the cloud dynamics determines the variability, then the luminosity of the emission appears to be independent of the mass
of the cloud:
\be \label{eq:L_g_sij}
L_{\gamma}\simeq {c P_0\pi R_{\rm c}^2 } {\xi\d_{\rm j}^4\over 4\G_{\rm j}^2}\,.
\ee
Since $L_{\rm j}>c P_0\pi R_{\rm c}^2$, the above  equation allows us to obtain a
lower limit on the jet luminosity required for the operation of the star-jet interaction scenario:
\be \label{eq:L_j_sij2}
L_{\rm j} > 0.4\zeta^2\d_{\rm j,1}^{-2}{L_{\gamma}}\xi_{-1}^{-1}\,,
\ee
or
\be \label{eq:L_j_sij}
L_{\rm j} > 0.025 \left({1+\alpha^2}\right)^4\G_{\rm j,1}^{-2}{L_{\gamma}}\xi_{-1}^{-1}\,,
\ee
which is a factor of $4/\Phi$ larger than the estimate for the jet-in-jet scenario (see Eq.~\eqref{eq:j_in_j_lum_total}).

Eqs.~\eqref{eq:t_var_sij}~and~\eqref{eq:energy_sij} contain four parameters $P_0$, $\G_{\rm j}$ (we treat $\d_{\rm j}$ as a
related parameter), $M_{\rm c}$, and $R_{\rm c}$, and therefore formally allow a solution even if two of these
parameters are fixed. For example, for given properties of the jet (i.e., for specific values of the parameters $\G_{\rm j}$ and $P_0$)
and the parameters characterizing the flare (the total energy and the variability), the characteristics of the cloud can be determined as
\be\label{eq:sol_sij}
M_{\rm c}c^2={E_\gamma\over \xi\d_{\rm j}^3}\,,\quad
\pi R_{\rm c}^2={4\zeta^2\over \d_{\rm j}^2}{E_\gamma\over \xi t_{\rm var} cP_0}\,.
\ee
However, the determined parameters of the cloud may not necessary be physical, and their feasibility should be examined by dynamical estimates.

The first dynamical limitation is related to the ability of a cloud to penetrate the jet and become involved in the jet
motion.  According to the estimates given by \cite{babkk10} and \cite{kbbad13}, for the typical jet parameters these constraints do not
impose any strong limitations. The heaviest blobs that can be accelerated by a jet with luminosity $10^{43}\rm\,erg\,s^{-1}$ 
can result in flares with a total energy release of $10^{54}\rm \,erg$. 

If the cloud is light enough to be caught by the jet, then one should consider two main processes: the cloud expansion,
and its acceleration. At the initial stage, the cloud cross-section is not sufficiently large to provide its acceleration to
relativistic velocities. On the other hand, the intense jet-cloud interaction at this stage leads to a rapid heating and expansion of the
cloud.  The cloud size-doubling time can be estimated as 
\be
t_{\rm exp} \approx A 
\left(\frac{M_{\rm c}}{ \gamma_g R_{\rm c} P_{\rm j}}\right)^{1/2}\,.
\label{tdoubl}
\ee
where $\gamma_g=4/3$ is the adiabatic index and $A$ is a constant of about a few \citep{gmr00,nmk06,phf10,bpb12}. When the
size of the cloud becomes large enough for acceleration to relativistic energies, the intensity of the jet-cloud interaction fades away, 
and the cloud expansion proceeds in the linear regime. Since the time scale for acceleration to relativistic velocity is 
\be \label{eq:time_acc}
t_{\rm ac} \simeq {M_{\rm c}c^2\over \pi R_{\rm c}^2 cP_{\rm j}}\,,
\ee
the size of the cloud relevant for the flare generation can be obtained by balancing Eqs.~\eqref{tdoubl}~and~\eqref{eq:time_acc}:
\be\label{eq:cloud_size}
R_{\rm c} = A_{\rm exp}\left(\frac{M_{\rm c}c^2}{P_{\rm j}}{\gamma_{\rm g}\over \pi^2 A^2}\right)^{1/3}\,.
\ee
Here the constant $A_{\rm exp}$ accounts for the cloud expansion in the linear regime.

The dynamical limitation given by Eq.~\eqref{eq:cloud_size} together with Eq.~\eqref{eq:sol_sij}
allows determination of the jet ram pressure:
\be
P_{\rm j} = \frac{\pi A^4}{\xi \gamma_{\rm g}^2 A_{\rm exp}^6}\left({2\zeta}\right)^6\frac{E_{\gamma}}{t_{\rm var}^3 c^3} \,.
\label{pjram}
\ee
The actual value of the coefficient in the above equation, in particular the value of $A_{\rm exp}$,
can be revealed only through the numerical simulations given the complexity of the jet-cloud interaction. However, if one
assumes that the expansion proceeds very efficiently, i.e., the cloud size achieves a value close to the light-crossing
limit, $R_{\rm c}\simeq\d_{\rm j}t_{\rm var}c$, then the expression for the jet ram pressure becomes
\be
P_{\rm j} =\left({2\zeta\over \d_{\rm j}^2}\right)^2\frac{E_{\gamma}}{\pi \xi t_{\rm var}^3 c^3} \,.
\label{pjram_simple}
\ee

Since each flaring episode should correspond to specific jet parameters\footnote{We note, however, that across a
  magnetically driven jet one may expect strong gradients of the jet ram pressure \citep[see,
  e.g.,][]{bn06,kbvk07,kvkb09}).}, the above equation implies that the energy emitted in an individual peak of a flare
should be proportional to the cube of its duration: $E_\gamma\propto t_{\rm var}^3$ (or
$L_\gamma \propto t_{\rm var}^{2}$). Obviously, the study of individual peaks in a statistically meaningful way
requires a detailed light curve that can be obtained with future observations, in particular with CTA \citep[see, e.g.,][]{rta17}.

\begin{table*}[]\label{tab:flares}
\centering
\caption{Comparison of models for different sources}
\label{tab:models}
\begin{tabular}{lccccc}
\tableline
\tableline
   Source              & \ic\tablenotemark{a}     & \quad\m87\tablenotemark{b}      & \c454\tablenotemark{c}    & \cc\tablenotemark{d}      & \pks\tablenotemark{e}  \\[2pt]
\tableline
&&&&&\\[-5pt]
 $M_{\rm BH,8}$  & \quad 3 & \quad 60  & \quad 10 &\quad 5   & \qquad 10 \\
 $t_{\rm 5}$     & \quad 1 & \quad 175 & \quad 54 &\quad 8.4 & \qquad 0.6 \\
 $\tau_{0}$      & \quad 0.2 & \quad 2 & \quad 3  &\quad 1   & \qquad 0.04 \\
 $L_{\gamma},\rm\,erg\,s^{-1}$    & \quad $2\times10^{44}$ & \quad $10^{42}$  & \quad $2\times10^{50}$ &\quad $6\times10^{48}$ & \qquad $10^{47}$ \\
  $\Phi$         & \quad 0.1& \quad 0.3& \quad 0.7&\quad 0.3 & \qquad 0.7\\[2pt]
\tableline
&&&&&\\[-5pt]
 $\Gamma_{\rm j}$& \quad 10 & \quad 10 & \quad 20 &\quad 20  & \qquad 20 \\
$\Gamma_{\rm co}$& \quad 10 & \quad 10 & \quad 10 &\quad 10  & \qquad 10 \\
  $\alpha$       & \quad 2  & \quad 2  & \quad 0  &\quad 0   & \qquad 0 \\[2pt]
\tableline
&&&&&\\[-5pt]
$L_{\gamma}/L_{\gamma,ms}$ & \quad $10$ & \quad $5\times10^{-4}$ & \quad $3\times10^{5}$        &\quad $5\times10^{4}$        & \qquad $10^{4}$ \\
$L_{\rm j, jj}$ & \quad $10^{44}$ & \quad $10^{42}$        & \quad $2\times10^{47}$        &\quad $3\times10^{45}$        & \qquad $ 10^{44}$ \\
$L_{\rm j, cj}$ & \quad $3\times10^{45}$ & \quad $2\times10^{43}$ & \quad $10^{48}$ &\quad $4\times10^{46}$ & \qquad $6\times 10^{44}$ \\[2pt]
\tableline
\end{tabular}
\tablenotetext{1}{\citet{ic310_14}}
\tablenotetext{2}{\citet{gt09}, \citet{wz09}, \citet{m87_12}.}
\tablenotetext{3}{\citet{agile10}, \citet{fermi11_3C}}
\tablenotetext{4}{\citet{hnm15}}
\tablenotetext{5}{\citet{ah07pks}}
\tablecomments{$M_{\rm BH,8}=M_{\rm BH}/10^8 M_\sun$ is the SMBH mass, $t_5=t/300 $~s is the variability time, $\tau_0=tc/r_{g}$ is the 
nondimensional variability time in units of gravitation radius light-crossing time, $L_{\gamma}$ is the maximum luminosity in gamma-rays,  
$\Gamma_{\rm j}$ is the jet Lorentz factor, $\Gamma_{\rm co}$ is the Lorentz factor of the mini-jet,  $\alpha=\theta/\G_{\rm j}$ is the  
normalized viewing angle, $L_{\gamma,ms}$ is the upper limit of the gamma-ray luminosity for a magnetospheric model, 
$L_{\rm j, jj}$ is the minimal jet power for the jet-in-jet model, $L_{\rm j, cj}$ is the minimum jet power for cloud-in-jet model. }
\end{table*}

\subsection{Energetic Constraints for Detected Exceptional Flares}
So far, several super-fast gamma-ray flares have been detected in the VHE or HE regimes from different types of AGNs. The peculiarity of the
signal is related both to the duration of the flare and to the released energy. Below we consider several
cases that are summarized in Table~\ref{tab:models}.

\subsubsection{\pks }
The July 2006 flare of \pks is characterized by a very short variability of $180\rm \,s$ and the a intrinsic
VHE gamma-ray luminosity at the level of $10^{47}\rm\,erg\,s^{-1}$  \citep{ah07pks}.  

The mass of the central BH is estimated to be $M_{\rm BH,8}\simeq 10$ \citep[][and reference therein]{ah07pks}. 
Together with the short variability time,
this constrains the luminosity of (potential) gamma-ray flares produced by magnetosphere gaps at the level of
$L_{\gamma,\rm ms}<10^{43}\rm\,erg\,s^{-1}\,$  and thus excludes any magnetospheric origin of these flares.

\pks is a typical representative of high-energy peaked BL Lacs. It is expected that the jet is aligned along the line of sight:   
$\alpha\approx0$. For a typical value of the jet Lorentz factor, $\Gamma_{\rm j}=20$, the jet power required for the realization of the 
jet-in-jet scenario is 
\be
L_{\rm j,jj}>10^{44} \Phi_{-0.2}\G_{\rm j,1.3}^{-2} \xi_{-1}^{-1}\,.
\ee

{ It follows from the comparison of Eqs.~\eqref{eq:j_in_j_lum_total2} and \eqref{eq:L_j_sij} that the lower limit of the jet power 
in the cloud-in-jet scenario is higher by a factor $\sim4/\Phi$, i.e.,}
\be 
L_{\rm j, cj} > 6\times10^{44}\Gamma_{\rm j,1.3}^{-2}\xi_{-1}^{-1}\,\rm erg\,s^{-1}\,.
\ee
We note here that constraints imposed by the radiation mechanism enhance the required jet power by a factor of $\sim10$ for the external inverse 
Compton scenario and by $\sim10^2$ for proton synchrotron emission \citep[see][]{babkk10}, which exceeds the Eddington luminosity.

\subsubsection{\ic }

In 2012 November, the \magic collaboration detected a bright flare from \ic \citep{ic310_14}. The flare consisted of two sharp peaks with
a typical duration of $\sim5\rm\,min$. The measured spectra were hard, with a photon index $\lesssim2$, extending up to
$\sim 10\rm TeV$. The energy released during this event has been estimated to be at a level of $2\times10^{44}\rm\,erg\,s^{-1}$.

The mass of the BH powering activity of \ic has been estimated to be $M_{\ic}=\left(3_{-2}^{+4}\right)\times10^8M_\sun$
\citep{ic310_14}, i.e., the measured variability time scale is as short as 20\% of $\tau_0$.

According to the estimate provided by Eq.~\eqref{lgmm}, the luminosity of flares generated in the BH magnetosphere
depends weakly on the mass of the BH and is determined by the disk magnetization, the viewing angle, and the pair
multiplicity\footnote{Eq.~\eqref{lgmm} does not account for relativistic effects that should be small unless the
  gap is formed close to the horizon. However, if the vacuum gap is close to the horizon, then the gravitational redshift should make more robust the constraints imposed by the variability time.}. Since all these parameters are smaller than unity, from Eq.~\eqref{lgmm} we have
\be
L_{\gamma,\rm ms}<2\times10^{43}\rm\,erg\,s^{-1}
\ee
This upper limit is an order of magnitude below the required value \citep{ic310_14}. Thus, we conclude that the ultrafast flare detected 
from this source cannot have a magnetospheric origin.}

Assuming that mini-jets are distributed isotropically in the jet frame and that the detection of two pulses is not a
statistical fluctuation, one can estimate the true jet luminosity using Eq.~\eqref{eq:j_in_j_lum_total2}. For the
relevant flare parameters (i.e., \(t_{\rm var}=4.8\rm \, min\), \(L_{\gamma}=2\times10^{44}\rm \,erg\,s^{-1}\)) and
\(M_8=3\)
\be
L_{\rm j,jj}> 10^{44} \Phi_{-1}\left({1+\alpha^2\over5}\right)^4\G_{\rm j,1}^{-2} \xi_{-1}^{-1}\,.
\ee
{If the mini-jets are not distributed isotropically,  the requirement on the jet power can be a few orders of magnitude weaker; see Eq.~\eqref{eq:j_in_j_lum}.}

The cloud-in-jet scenario requires a higher jet luminosity; from Eq.~\eqref{eq:L_j_sij} it follows that 
\be 
L_{\rm j, cj} > 3\times 10^{45}\Gamma_{\rm j,1}^{-2}\left({1+\alpha^2\over5}\right)^4\xi_{-1}^{-1}\,\rm erg\,s^{-1}\,.
\ee
%

\subsubsection{\m87 }
In 2010, a bright flare has been recorded during a multiintrument campaign in the VHE energy band \citep{m87_12}. The
variability time during the VHE transient was about $0.6$~day and the flux level achieved $10^{42}\rm\, erg\,s^{-1}$. This source is
characterized by a large jet-viewing angle of $\theta_{\rm j}\approx 15^o$ and a Lorentz factor of about $\Gamma_{\rm j}\approx 7$ \citep{wz09}, 
and the SMBH mass is $\sim6\times10^9M_\sun$ \citep{gt09}. 

Given the heavy central BH and the relatively long duration of the VHE flare, which allows high values of the gap size, 
the energy constraint in the magnetosphere scenario is quite modest:
\be
L_{\gamma,\rm ms}<2\times10^{45}\rm\,erg\,s^{-1}\,.
\ee
\m87 might be an interesting
candidate for a detection of magnetosphere flares. 

For the flare parameters  (i.e., $t_{\rm var}=0.6\rm \, d$, $L_{\gamma}=10^{42}\rm \,erg\,s^{-1}$) and $M_8=60$,
Eq.~\eqref{eq:j_in_j_lum_total2} constrains the required jet true luminosity at the level
\be
L_{\rm j,jj}> 10^{42} \Phi_{-0.5}\left({1+\alpha^2\over5}\right)^4\G_{\rm j,1}^{-2} \xi_{-1}^{-1}\,.
\ee
{ On the other hand, the mulitwavelength properties of the gamma-ray flares detected from \m87 seem to be quite diverse,
with no detected robust counterparts at other wavelengths. Thus, if the VHE emission is produced by a single mini-jet, 
then a much weaker constraint, provided by Eq.~\eqref{eq:j_in_j_lum}, is applied. In this case, the variability detected with Cherenkov 
telescopes should correspond to the mini-jet variability, thus the mini-jet comoving size should be 
\be
\tilde{l}_{\rm em}=\Delta t c \Gamma_{\rm em}={2\Delta t c \Gamma_{\rm j}\Gamma_{\rm co}\over 1+\alpha^2}\sim10^{17}\rm \,cm\,,
\ee
which is about the jet cross-section at a parsec distance from the central BH. 
 We should also note that the typical spectra emitted by plasmoids  are dominated by synchrotron radiation, which seems to be inconsistent 
 with the multiwavelength observations of \m87. Moreover,  the peculiar light curve that has been detected with
\hess{} has not yet been explained in the framework of the jet-in-jet scenario.

Formally, for the parameters of the flare detected from \m87, the minimum jet luminosity required by the cloud-in-jet scenario is
\be 
L_{\rm j, cj} > 2\times10^{43}\Gamma_{\rm j,1}^{-2}\left({1+\alpha^2\over5}\right)^4\xi_{-1}^{-1}\,\rm erg\,s^{-1}\,.
\ee
However, it has been argued that the light curve and the VHE spectrum is best explained if the TeV is produced through p-p 
interactions induced by the jet collision with a dense cloud. In this case, the required jet power is about 
$L_{\rm j}\approx5\times10^{44}\rm\,erg\,s^{-1}$ \citep{bba12}.

\subsubsection{\c454 and \cc}
In 2010 November, an exceptionally bright flare was detected from  \c454{}  by \ag and \fer \citep{agile10,fermi11_3C}. 
The minumum detected variability time and gamma-ray 
luminosity were $4.5$~hours and $2\times 10^{50}$~erg~s$^{-1}$, respectively. 

Several similarly  bright flares were detected form \cc in the period from  2013 December to  2014 April \citep{hnm15}.
The GeV gamma-ray luminosity reached a level of $6\times 10^{48} \rm\,erg\,s^{-1}$, and the flux varied on a time scale of 0.7~hr. 

The magnetosphere gap luminosity is limited by $10^{45}\rm\,erg\,s^{-1}$ and $2\times10^{44}\rm\,erg\,s^{-1}$ for
\c454 and \cc, respectively.{Therefore the magnetospheric origin of these flares is excluded for both sources. }

\cc and \c454 are distant quasars, thus it is safe to fix $\alpha=0$ for both cases. By adopting a standard jet Lorentz factor, 
$\Gamma_{\rm j}=20$, one can obtain the jet luminosity required for the realization of the jet-in-jet scenario:
\be
L_{\rm j,jj}> 10^{47} \Phi_{-0.2}\G_{\rm j,1.3}^{-2} \xi_{-1}^{-1}\,
\ee
for \c454, and 
\be
L_{\rm j,jj}> 3\times10^{45} \Phi_{-0.5}\G_{\rm j,1.3}^{-2} \xi_{-1}^{-1}\,
\ee
for \cc. Both these estimates appears to be below the corresponding Eddington luminosity limits of $1.3\times10^{47}\rm\,erg\,s^{-1}$ and $6\times10^{46}\rm\,erg\,s^{-1}$  for $M_{\rm BH,8}=10$ and $M_{\rm BH,8}=5$ in \c454 and \cc, respectively.

The star-in-jet scenario requires higher jet luminosities, which seem to exceed the Eddington limit for \c454. Namely, one obtains 
\be 
L_{\rm j, cj} > 10^{48}\Gamma_{\rm j,1.3}^{-2}\xi_{-1}^{-1}\,\rm erg\,s^{-1}\,.
\ee
This value agrees with the estimate $L_{\rm j, cj}\approx 10^{49} \rm\, erg\,s^{-1}$ that is obtained within a more accurate
model that provides also provides an interpretation for the so-called plateau phase and the spectrum \citep{kbbad13}. Such
a high luminosity of the jet, $L_{\rm j}\approx 0.1 L_{\gamma}$, has also been  obtained in the framework of the one-zone
external Compton model of \cite{bgf11}.

For \cc, the lower limit on the jet luminosity is
\be 
L_{\rm j, cj} > 4\times10^{46}\Gamma_{\rm j,1.3}^{-2}\xi_{-1}^{-1}\,\rm erg\,s^{-1}\,,
\ee
which is close to the Eddington limit. A similar estimate was obtained by \cite{hnm15}.

\section{Discussion and Conclusions}
\label{sc:diss}

\subsection{Gamma-ray Flare Detected from IC 310}
\ic is a radio galaxy with redshift \(z\simeq0.0189\) \citep{uzc}. Radio observations revealed an extended jet with a viewing 
angle of \(\theta_{\rm j}< 30^\circ\) \citep[see, e.g.,][]{ker12}. Arguments based on the absence of the a contra-jet and assuming that 
the true length of the jet is smaller than \(1\rm Mpc\) allowed to further constrain the viewing angle 
\(10^\circ<\theta_{\rm j}<20^\circ\) \citep[for details see][]{ic310_14}. Finally, observations of superluminal motion allowed us
to constraint the Doppler factor to \(\d_{\rm j}\sim5\).  The radio luminosity has been estimated to be at the level of 
\(L_{\ic,radio}\simeq10^{41}\rm \, erg\,s^{-1}\) \citep{SijBru98}. This implies a minimum energy in relativistic electrons of 
\(5.61\times10^{57}\rm \,erg\), thus the power required for the supply of emitting electrons yields 
\(L_{\ic,e}\simeq2\times10^{42}\rm \, erg\,s^{-1}\), and the total jet luminosity can be estimated as 
\(L_{\ic}\simeq10^{43}\rm\,erg\,s^{-1}\) \citep[see][]{ic310_14}. We note that these estimates represent values for the 
{\it minimum} required energetics {\it averaged} over \(10^8\)~years.

In 2012 November, \magic detected a bright flare from \ic \citep{ic310_14}. The flare consisted of two sharp peaks with a typical 
duration of $\sim5\rm\,min$. The measured spectrum was hard, with a photon index of$\lesssim2$, extending up to $\sim 10\rm\, TeV$. 
The energy released during that event has been estimated to be at a level of $2\times10^{44}\rm\,erg\,s^{-1}$.

The mass of the BH powering activity of \ic has been determined to be \(M_{\ic}=\left(3_{-2}^{+4}\right)\times10^8M_\sun\) 
\citep{ic310_14}, i.e., the measured variability time scale is as short as 20\% of $\tau_0$; therefore one can expect 
a realization of some unconventional mechanism for VHE emission production. \citet{ic310_14} have considered possible scenarios 
(see Sect.~\ref{sc:energy}) for the flare production and found that jet-in-jet and star-in-jet interaction models face certain difficulties. 
Based on this, one concluded that the magnetosphere origin remains the only possible option, and no further verification of that scenario 
has been provided.

\citet{hp16} have performed simulations of the gamma-ray spectrum
produced in a stationary gap for different accretion rates and
concluded that the emission generated in the vacuum gap could closely
reproduce the spectral properties of the TeV emission detected during
the flare. The strength of the magnetic field has been fixed at
  a level of $10^4\rm\, G$. \citet{hp16} have emphasized that
  the generation of such a strong magnetic field requires an accretion rate
  exceeding the values compatible with the existence 
  of vacuum gaps by a factor of \(100\). In addition, we note that if the thickness of the gap
is determined by the variability time, $h\simeq t_{\rm var}c$, this
scenario requires an even higher efficiency \citep[by a factor of
$\sim20$, since the best-fit parameters for the magnetospheric scenario
require $\dot{m}\sim5\times10^{-6}$ and consequently
$h_{\rm gap}\sim5r_{\rm g}(\sim 25t_{\rm var}c)$, see Figures 8 and 17
of][]{hp16}.

Since one does not expect any significant focusing or enhancement of the emission produced in the magnetosphere, 
the measured energy should correspond to the real energetics of the processes responsible for the emission generation. 
Thus, the feasibility of generating such a powerful flare in the vacuum gap is closely related to the general efficiency 
of processes taking place in BH magnetosphere. Currently, the BZ mechanism \citep{ruffini75, lovelace76, BZ77} represents 
the most prominent energy extraction mechanism that can operate in the BH magnetosphere. The efficiency of this mechanism is 
determined by the strength of the magnetic field that is accumulated at the BH horizon, which in turn is determined by the accretion rate. 
\citet[][]{hp16} argued that the efficiency of the BZ mechanism can be very high, up to a level of 900\%, as compared to 
the accretion rate $\dot{M}c^2$. 
This assumption is based on 2D simulations presented by \citet{mtb12}. However, 3D simulations for a similar setup presented in 
the same paper reveal a significantly lower efficiency, $\sim300$\%. We note that Eq.~(\ref{lgam_m}) with $\alpha_{\rm ss}=0.1$ 
corresponds to an efficiency of 250\%, which is very close to the results of the 3D simulations by \citet{mtb12}.

According to the estimate provided by Eq.~\eqref{lgmm}, the possible luminosity of flares generated in BH magnetosphere depends 
very weakly on the mass of the BH and is determined by disk magnetization, viewing angle, and pair multiplicity\footnote{Eq.~\eqref{lgmm} 
does not account for relativistic effects, which should be small unless the gap is formed close to the horizon. If the vacuum gap is close 
to the horizon, then gravitational redshift should even strengthen the constraints imposed by the variability time.}. Since all these 
parameters are smaller than one, then the numerical coefficient in Eq.~\eqref{lgmm} can be taken as a strict upper limit for the flare 
luminosity for the given variability time. This upper limit appears to be approximately an order of magnitude below the value measured with 
\magic \citep{ic310_14}. Thus, we conclude that it seems very unfeasible that these processes are indeed behind the bright flaring activity 
recorded from \ic \citep[which is also consistent with modeling presented by][]{hpc16}.  

Our simplified analysis does
not allow us to robustly rule out two other scenarios for the flare production in \ic. If one adopts the minimum averaged jet 
luminosity as a reasonable constraint for the present jet luminosity \citep[as done in][]{ic310_14}, then both scenarios 
formally do not allow reproducing the observed properties. However, if one assumes that there is some anisotropy in the mini-jet distribution, 
then jet-in-jet model provides an energetically feasible scenario. We note that the spectral energy distributions currently obtained for 
the jet-in-jet models feature the dominant excess in the UV band \citep{pgs16}, which is not consistent with the observations from \ic. 
This might be either a fundamental constraint or just a systematic underestimation of the inverse Compton contribution that is due 
to the small scale of the simulations. 
Thus, it seems that more detailed large-scale simulations are required to verify the applicability of the jet-in-jet scenario for \ic. 
On the other hand, if the present-day jet luminosity is significantly higher than the averaged value, the star-in-jet scenario may also meet 
the energetic requirements.

\subsection{Comparison of Scenarios for Ultrafast Variability}
In this paper we considered three scenarios for the production of ultrafast AGN flares with variability times shorter than the Kerr 
radius light-crossing time: gamma-ray emission of gaps in the SMBH magnetosphere \citep{NerAha07,lr11}, the jet-in-jet 
realization \citep{gub09}, and the emission caused by penetration of  external dense clouds \citep{babkk10}.

The production of gamma rays in the BH magnetosphere has several unique properties. In particular, this scenario can be invoked 
to explain emission from off-axis AGNs and orphan gamma-ray flares. On the other hand, the luminosity of the magnetospheric gap has a robust upper 
limit that depends weakly on the SMBH mass. Moreover, the magnetospheric emission is not enhanced by the Doppler-boosting effect, 
and this seems to be crucial for explaining short flares from distant AGN. On the other hand, some nearby SMBHs\citep{lr11}, e.g., 
the Sagittarius A star or \m87, 
might be very promising candidates to produce gamma-ray flares \citep[see, however,][for the discussion of gamma-gamma 
attenuation in magnetosphere]{lyw09,lr11,cyl12}.

In general terms, there can be little doubt that the nonthermal
radiation of powerful AGN is related, in one way or another, to
relativistic jets. The ultrafast gamma-ray flares might be linked to the formation of relativistically moving features
(plasmoids or mini-jets) inside the major outflow, the jet originating
from the central black hole.  Depending on the orientation of the
mini-jets to the jet axis, the radiation of the mini-jet can be
focused within the jet cone or outside.  This scenario has been
suggested to interpret the variable emission from AGN
\citep{gub09,gub10}. It has been shown that under certain
  conditions, magnetic field reconnection can result in the formation of
relativistic outflows \citep{lub05,sgp16}. {We note, however, that
  formation of a relativistic outflow is not an indispensable feature
  of reconnection. Thus, ejection of relativisitcally moving plasmoids
  may require a specific configuration of the magnetic
  field. Independently, to form outflows with large Lorentz
  factors, \(\G_{\rm co}\geq10\),
  an initial configuration with high magnetization,
  \(\sigma\simeq \G_{\rm co}^2\geq100\),
  is required. Such a high magnetization of the jet at the flare
  production site requires an even higher initial jet magnetization,
  \(\sigma_{\rm init}\gg10^3\).
  Jets with such a high magnetization should have an extremely low
  mass load, which seems to be inconsistent with the properties of AGN jets
  at large distances \citep[see, however,][and references
  therein]{k94,sp06,abr13}.

Finally, the SED of the emission produced by plasmoids formed at reconnection contains a dominating synchrotron component that
peaks in the UV energy band \citep{pgs16}.  This feature is not consistent with the SEDs obtained from AGNs during the
ultrafast flares. The presence of a guiding magnetic field can significantly enhance the magnetization of
plasmoids, resulting in a further enhancement of the synchrotron component and perhaps in the extension of the
synchrotron component to the gamma-ray band. The examination of this scenario requires detailed modeling, since the 
guiding filed also impacts the Lorentz factor of plasmoids.}

The jet-in-jet scenario quantitatively implies a modest requirement for the jet intrinsic luminosity, however; it can be even
further relaxed if one assumes that the mini-jets are not distributed isotropically in the major jet comoving
frame. Such an anisotropy can be realized, for example, by focusing the outflow along the direction of the reconnecting
magnetic field.

An important issue to realize the jet-in-jet scenario
what is the triggering mechanism for the reconnection. If the jet is
launched by the BZ mechanism, it is expected to be
magnetically dominated at the initial stage, thus reconnection is a
thermodynamicaly favored process. However, the reasoning based on
equipartition arguments, without any particular energy transfer
mechanism, can hardly be valid. The time-scale of such thermodynamic
processes may be enormous; thus they might be irrelevant for
astrophysical jets. In recent years, significant progress has been
achieved in PIC simulations for the reconnection in magnetic field
configurations with alternating polarities. Such configurations should
naturally appear in the pulsar outflows close to the current
sheet. However, it is less obvious how such regions would form in AGN
jets. Several scenarios can be considered.  The first is a change
in the magnetic field polarity in the jet caused by a change in
magnetic field polarity in the accretion disk and, consequently, in
the BH magnetosphere \citep{bb11,mu12}. However, such a change takes a
long time, of about \(\sim10^3r_{g}/c\),
and it is hard to expect ultrafast variability caused by such a
configuration of the magnetic field. An arrangement with
alternating magnetic field polarities can also be a result of a growth of
MHD instabilities in the jet \citep[see, e.g.,][]{dtg16}. However, an
intense instability growth leads to the a flow disruption on a scale of
several dynamical lengths \citep{pk15}. So it is hard to obtain an
intensive reconnection event close to the base of a jet that extends
a significant distance beyond the reconnection region. Finally, the
reconnection can be caused by a sudden compression and mixing of a
small part of the jet, which, for example, can be due to an external
obstacles in the jet.  In such a case, a short but intensive local
reconnection episode may occur without disrupting the entire
flow. This specific case represents an interesting synergy of two
models: the formation of a relativistic mini-jet by reconnection of the
magnetic field triggered by a star in the jet. The feasibility of this
scenario needs to be tested with detailed numerical simulations.

The star-in-jet scenario, the third possibility considered in the paper, requires significantly higher jet
luminosity than the jet-in-jet scenario. In many cases, the jet luminosity, needed to realize the star-in-jet
scenario, exceeds the Eddington limit. It was also shown that some details of the GeV light curve obtained from \c454 with \fer, e.g., the
plateau phase, can be readily interpreted in the framework of the star-in-jet scenario \citep{kbbad13}. It is also
important to note that the emission produced by the interaction of a cloud with the AGN jet should be characterized by a
universal relation between the luminosity and the duration of individual peaks of the flare: $L^{1/2}\propto \Delta t$.
To verify this relation observationally, a high photon statistics is required, which may possibly be achieved with future
observations with CTA.

\appendix

\section{Calculation of the Pair Multiplicity from a Radiatively Inefficient Accretion Flow}
\label{app:npm}
\cite{lr11} have estimated the density of electron-positron pairs produced by photon-photon annihilation in a radiatively
inefficient accretion flow \citep[RIAF,][]{ny95}. For the sake of consistency, a similar consideration is present below,
but for a Kerr BH and explicitly accounting for the nondimensional viscosity parameter $\alpha_{\rm ss}$ and radiation
efficiency off an accretion flow $\eta$. Identically to \cite{lr11}, we relie on the solution obtained by \citet{ny95},
assuming that the advection parameter is small: $1-f\ll1$ \citep[with notations of][]{ny95}.  In particular, the radial
velocity of the accretion flow is taken as 
\be 
v_r=\frac{3\alpha_{\rm ss}}{5}\left(\frac{G M_{\rm bh}}{r}\right)^{1/2}\,.  
\ee
The ion density $n_i$ can be estimated as 
\be 
n_i(r)=\frac{\dot{M}}{4\pi rH m_p v_r} = \frac{5\sqrt{10}}{6}
\frac{\dot{m}}{\eta \alpha_{ss}} \frac{(G M_{\rm bh})^{1/2}}{c \sigma_{\rm T} r^{3/2} }\,,
\ee
given the revealed height of the accretion disk, $H\simeq c_{\rm s}/\Omega_{\rm k}$ \citep[for details and notations see][]{ny94}.

The total cooling rate of the ion-electron plasma can be estimated \citep{ny95b,lr11} as 
\be
q_{ff}= q_{ee}+q_{ei}\approx 10^{-21} n_e^2\theta_e \mbox{ erg s}^{-1} \mbox{ cm}^{-3}
\ee
for relativistic electron temperature $\theta_e = kT_e/m_e c^2\gtrsim1$. Since the density of the ions exceeds the GJ density,
the pair production does not provide any sensible contribution to the disk electron density in configurations allowing
the existence of vacuum gaps, thus in what follows we assume the number densities of electrons and ions to be equal,
$n_i = n_e$.

For such a hot electron plasma the emission appears in the MeV energy band, and the luminosity of the inner part of the accretion 
flow is \citep{lr11}
\be\label{eq:l_ff}
L_{ff} \approx \int_{r_{g}}^{2r_{g}} 2\pi r^2 q_{ff} dr\,.
\ee
A lower limit on the number density of these MeV photons is 
\be
n_{\gamma}\approx \frac{L_{ff}}{4\pi c (2r_{g})^2 e_{\gamma}}\approx \frac{0.7 q_{ff} r_g^3}{c r_g^2 e_{\gamma}}\sim
\frac{ 10^{9} \dot{m}^2}{\alpha_{ss}^2\eta^2 M_{8}} \mbox{ cm}^{-3},
\ee
where $e_{\gamma}=3\theta_e m_e c^2$. The production rate of $e^{\pm}$ pairs inside the magnetosphere due to
$\gamma\gamma$-annihilation is approximately $\sigma_{\gamma\gamma} n_\gamma^2c(4\pi/3)(2r_g)^3$, where
$\sigma_{\gamma\gamma}\approx\sigma_{\rm T}/5$ is the cross-section of two-photon pair production.  In steady state, this
rate is balanced by the escape rate $\sim 4\pi c (2r_g)^2 n_{\pm}$, allowung us to estimate the 
\be 
n_\pm\gtrsim 10^6 \frac{\dot{m}^4}{\eta^4\alpha_{\rm ss}^4 M_{ 8}}
\mbox{ cm}^{-3}.  
\ee
The GJ density (ignoring its polar angle dependence) is determined by the accretion rate, $\dot{m}$, via Eqs.~(\ref{bpp})~and~(\ref{eq:omega}):
\be 
n_{\rm GJ}=\frac{\Omega B}{2\pi e c} = 0.4 \left(\frac{\beta_m \dot{m}}{\eta\alpha_{\rm ss} M_{\rm
      bh,8}^3}\right)^{1/2} \mbox{ cm}^{-3}.  
\ee 
Thus, the multiplicity parameter is 
\be 
\kappa \gtrsim \frac{n_\pm}{n_{\rm GJ}} \approx 6\times 10^6 \frac{\dot{m}^{7/2} M_{\rm
    bh,8}^{1/2}}{\eta^{7/2}\alpha_{\rm ss}^{7/2}\beta_m^{1/2}}\,.
\ee 
The condition $\kappa<1$ place upper limit on the accretion
rate as 
\be 
\dot{m}\lesssim 10^{-2} \frac{\eta\alpha_{\rm ss}\beta_m^{1/7}}{M_{8}^{1/7}}.  
\ee

{ Eq. \eqref{eq:l_ff} allows an estimation of the characteristic inverse Compton cooling time of electrons in the photon field provided by the disk. For an electron with energy \(E=10^6m_ec^2\gamma_6\)}
\be
\lambda_{\rm IC}= \frac3{4\sigma_t}{mc^2\over  n_\gamma e_\gamma 10^6\gamma_6 }\sim {\alpha_{\rm ss}^2\eta^2M_8\over \theta_e}{{4\times10^{8}\,\rm cm} \over \gamma_6 \dot{m}^2}\gtrsim {0.2 r_{\rm g}\over \gamma_6}{M_8^{2/7}\over \beta_{m}^{2/7}\theta_e}\,.
\ee
\section{Small Angle Limit}
\label{app:sal}
It is believed that the emission from AGNs is mostly detected by observers located within the jet-beaming cone
$\theta\leq\G_{\rm j}^{-1}$, but there are also some examples when the flares are detected from off-axis radio
galaxies, e.g., \m87 and \ic. Sometimes it is convenient to measure the viewing angle in the jet-opening units:
\be
\theta=\alpha\G_{\rm j}^{-1}\,.
\ee
In particular, this parameter gives a simple relation between the jet Doppler factor and the Lorentz factor:
\be
\d_{\rm j}={1\over \G_{\rm j}(1-\beta_{\rm j}\cos\theta)}={2\G_{\rm j}\over1+\alpha^2}\,,
\ee
which is valid for $\alpha\ll\G_{\rm j}$.

If the production region moves relativistically in the jet, then its Lorentz factor with respect to the observer is \citep[see, e.g.,][]{gub09} 
\be
\G_{\rm em}=\G_{\rm j}\G_{\rm co}(1+\beta_{\rm j}\beta_{\rm co}\cos\theta')\,,
\ee
where $\theta'$ is the angle between the jet velocity and the outflow velocity. 
This angle is related to the angle in the observer frame via the aberration formula,
\be
\tan\theta={\beta_{\rm co}\sin\theta'\over \G_{\rm j}(\beta_{\rm co}\cos\theta'+\beta_{\rm j})}\,.
\ee
The latter equation allows us to express the angle in the comoving reference frame
\be
\cos\theta'={-\alpha^2\frac{\beta_{\rm j}}{\beta_{\rm co}}\pm\sqrt{\alpha^4\frac{\beta_{\rm j}^2}{\beta_{\rm co}^2}+\left({1-\alpha^2\frac{\beta_{\rm j}^2}{\beta_{\rm co}^2}}\right)(1+\alpha^2)}\over 1+\alpha^2}\,.
\ee
One should account for the kinematic constraints that naturally appear in the above equation: $|\cos\theta'|\leq1$. Taking the solution with 
the $+$ sign (which corresponds to a stronger enhancement in the region where two solutions are allowed) in the limit $\G_{\rm j,co}\gg1$, 
one obtains
\be
\beta_{\rm co}\cos\theta'\simeq\frac{1-\alpha^2}{1+\alpha^2}-\frac1{2\G_{\rm co}^2}\,,
\ee
and consequently, the emitter Lorentz factor is
\be
\G_{\rm em}\simeq{2\G_{\rm j}\G_{\rm co}\over 1+\alpha^2}\,.
\ee

\section*{Acknowledgments}
The authors are grateful  to the anonymous referee for  the insightful comments and helpful suggestions. 
We would like to thank Andrey Timokhin and Frank Rieger for productive discussions. The authors appreciate the support by the Russian
Science Foundation under grant 16-12-10443.  D.K. acknowledges financial support by a grant-in-aid for Scientific
Research (KAKENHI, No. 24105007-1) from the Ministry of Education, Culture, Sports, Science and Technology of Japan
(MEXT). M.B. acknowledges partial  support  by the JSPS (Japan Society for the Promotion of Science):
No.2503786, 25610056, 26287056, 26800159. M.B. also acknowledges MEXT: No.26105521 and for partial support 
by NSF  grant AST-1306672 and DoE grant DE-SC0016369.

\bibliographystyle{apj} 
\bibliography{star_in_jet}

\begin{thebibliography}{87}
\expandafter\ifx\csname natexlab\endcsname\relax\def\natexlab#1{#1}\fi

\bibitem[{{Abdo} {et~al.}(2011){Abdo}, {Ackermann}, {Ajello}, {Allafort},
  {Baldini}, {Ballet}, {Barbiellini}, {Bastieri}, {Bellazzini}, {Berenji},
  {Blandford}, {Bloom}, {Bonamente}, {Borgland}, {Bouvier}, {Bregeon},
  {Brigida}, {Bruel}, {Buehler}, {Buson}, {Caliandro}, {Cameron}, {Caraveo},
  {Casandjian}, {Cavazzuti}, {Cecchi}, {Charles}, {Chekhtman}, {Cheung},
  {Chiang}, {Ciprini}, {Claus}, {Conrad}, {Cutini}, {D'Ammando}, {de Angelis},
  {de Palma}, {Dermer}, {Digel}, {Silva}, {Drell}, {Dubois}, {Dumora},
  {Escande}, {Favuzzi}, {Fegan}, {Ferrara}, {Fortin}, {Fukazawa}, {Fusco},
  {Gargano}, {Gasparrini}, {Gehrels}, { Germani}, {Giglietto}, {Giommi},
  {Giordano}, {Giroletti}, {Glanzman}, {Godfrey}, {Grenier}, {Grove},
  {Guiriec}, {Hadasch}, {Hayashida}, {Hays}, {Horan}, {Itoh},
  {J{\'o}hannesson}, {Johnson}, {Kamae}, {Katagiri}, {Kataoka},
  {Kn{\"o}dlseder}, {Kuss}, {Lande}, {Larsson}, {Latronico}, {Lee}, {Longo},
  {Loparco}, {Lott}, {Lovellette}, {Lubrano}, {Madejski}, {Makeev},
  {Mazziotta}, {McConville}, {McEnery}, {Michelson}, {Mitthumsiri}, {Mizuno},
  {Moiseev}, {Monte}, {Monzani}, {Morselli}, {Moskalenko}, {Murgia},
  {Naumann-Godo}, {Nishino}, {Nolan}, {Norris}, {Nuss}, {Ohsugi}, {Okumura},
  {Orlando}, {Ormes}, {Paneque}, {Pelassa}, {Pesce-Rollins}, {Pierbattista},
  {Piron}, {Porter}, {Rain{\`o}}, {Rando}, {Razzaque}, {Reimer}, {Reimer},
  {Ritz}, {Roth}, {Sadrozinski}, {Sanchez}, {Scargle}, {Schalk}, {Sgr{\`o}},
  {Siskind}, {Smith}, {Spandre}, {Spinelli}, {Strickman}, {Takahashi},
  {Takahashi}, {Tanaka}, {Tanaka}, {Thayer}, {Thayer}, {Thompson}, {Tibaldo},
  {Torres}, {Tosti}, {Tramacere}, {Troja}, {Vandenbroucke}, {Vasileiou},
  {Vianello}, {Vilchez}, {Vitale}, {Waite}, {Wang}, {Winer}, {Wood}, {Yang}, \&
  {Ziegler}}]{fermi11_3C}
{Abdo}, A.~A., {et~al.} 2011, \apjl, 733, L26

\bibitem[{{Abramowicz} {et~al.}(1988){Abramowicz}, {Czerny}, {Lasota}, \&
  {Szuszkiewicz}}]{acl88}
{Abramowicz}, M.~A., {Czerny}, B., {Lasota}, J.~P., \& {Szuszkiewicz}, E. 1988,
  \apj, 332, 646

\bibitem[{{Abramowski} {et~al.}(2012){Abramowski}, {Acero}, {Aharonian},
  {Akhperjanian}, {Anton}, {Balzer}, {Barnacka}, {Barres de Almeida},
  {Becherini}, {Becker}, \& et~al.}]{m87_12}
{Abramowski}, A., {et~al.} 2012, \apj, 746, 151

\bibitem[{{Ackermann} {et~al.}(2016){Ackermann}, {Anantua}, {Asano}, {Baldini},
  {Barbiellini}, {Bastieri}, {Becerra Gonzalez}, {Bellazzini}, {Bissaldi},
  {Blandford}, {Bloom}, {Bonino}, {Bottacini}, {Bruel}, {Buehler}, {Caliandro},
  {Cameron}, {Caragiulo}, {Caraveo}, {Cavazzuti}, {Cecchi}, {Cheung}, {Chiang},
  {Chiaro}, {Ciprini}, {Cohen-Tanugi}, {Costanza}, {Cutini}, F.~D'Ammando, {de
  Palma}, {Desiante}, {Digel}, {Di Lalla}, {Di Mauro}, {Di Venere}, {Drell},
  {Favuzzi}, {Fegan}, {Ferrara}, {Fukazawa}, {Funk}, {Fusco}, {Gargano},
  {Gasparrini}, {Giglietto}, {Giordano}, {Giroletti}, {Grenier}, {Guillemot},
  {Guiriec}, {Hayashida}, {Hays}, {Horan}, {J{\'o}hannesson}, {Kensei},
  {Kocevski}, {Kuss}, {La Mura}, {Larsson}, {Latronico}, {Li}, {Longo},
  {Loparco}, {Lott}, {Lovellette}, {Lubrano}, {Madejski}, {Magill}, {Maldera},
  {Manfreda}, {Mayer}, {Mazziotta}, {Michelson}, {Mirabal}, {Mizuno},
  {Monzani}, {Morselli}, {Moskalenko}, {Nalewajko}, {Negro}, {Nuss}, {Ohsugi},
  {Orlando}, {Paneque}, {Perkins}, {Pesce-Rollins}, {Piron}, {Pivato},
  {Porter}, {Principe}, {Rando}, {Razzano}, {Razzaque}, {Reimer}, {Scargle},
  {Sgr{\`o}}, {Sikora}, {Simone}, {Siskind}, {Spada}, {Spinelli}, {Stawarz},
  {Thayer}, {Thompson}, {Torres}, {Troja}, {Uchiyama}, {Yuan}, \&
  {Zimmer}}]{aaa16}
{Ackermann}, M., {et~al.} 2016, \apjl, 824, L20

\bibitem[{{Aharonian} {et~al.}(2007){Aharonian}, {Akhperjanian}, {Bazer-Bachi},
  {Behera}, {Beilicke}, {Benbow}, {Berge}, {Bernl{\"o}hr}, {Boisson}, {Bolz},
  {Borrel}, {Boutelier}, {Braun}, {Brion}, {Brown}, {B{\"u}hler},
  {B{\"u}sching}, {Bulik}, {Carrigan}, {Chadwick}, {Clapson}, {Chounet},
  {Coignet}, {Cornils}, {Costamante}, {Degrange}, {Dickinson},
  {Djannati-Ata{\"i}}, {Domainko}, {Drury}, {Dubus}, {Dyks}, {Egberts},
  {Emmanoulopoulos}, {Espigat}, {Farnier}, {Feinstein}, {Fiasson},
  {F{\"o}rster}, {Fontaine}, {Funk}, {Funk}, {F{\"u}{\ss}ling}, {Gallant},
  {Giebels}, {Glicenstein}, {Gl{\"u}ck}, {Goret}, {Hadjichristidis}, {Hauser},
  {Hauser}, {Heinzelmann}, {Henri}, {Hermann}, {Hinton}, {Hoffmann}, {Hofmann},
  {Holleran}, {Hoppe}, {Horns}, {Jacholkowska}, {de Jager}, {Kendziorra},
  {Kerschhaggl}, {Kh{\'e}lifi}, {Komin}, {Kosack}, {Lamanna}, {Latham}, {Le
  Gallou}, {Lemi{\`e}re}, {Lemoine-Goumard}, {Lenain}, {Lohse}, {Martin},
  {Martineau-Huynh}, {Marcowith}, {Masterson}, {Maurin}, {McComb}, {Moderski},
  {Moulin}, {de Naurois}, {Nedbal}, {Nolan}, {Olive}, {Orford}, {Osborne},
  {Ostrowski}, {Panter}, {Pedaletti}, {Pelletier}, {Petrucci}, {Pita},
  {P{\"u}hlhofer}, {Punch}, {Ranchon}, {Raubenheimer}, {Raue}, {Rayner},
  {Renaud}, {Ripken}, {Rob}, {Rolland}, { Rosier-Lees}, {Rowell}, {Rudak},
  {Ruppel}, {Sahakian}, {Santangelo}, {Saug{\'e}}, {Schlenker}, {Schlickeiser},
  {Schr{\"o}der}, {Schwanke}, {Schwarzburg}, {Schwemmer}, {Shalchi}, {Sol},
  {Spangler}, {Stawarz}, {Steenkamp}, {Stegmann}, {Superina}, {Tam},
  {Tavernet}, {Terrier}, {van Eldik}, {Vasileiadis}, {Venter}, {Vialle},
  {Vincent}, {Vivier}, {V{\"o}lk}, {Volpe}, {Wagner}, {Ward}, \&
  {Zdziarski}}]{ah07pks}
{Aharonian}, F., {et~al.} 2007, \apjl, 664, L71

\bibitem[{Albert {et~al.}(2007)Albert, Aliu, Anderhub, Antoranz, Armada,
  Baixeras, Barrio, Bartko, Bastieri, Becker, Bednarek, Berger, Bigongiari,
  Biland, Bock, Bordas, Bosch-Ramon, Bretz, Britvitch, Camara, Carmona,
  Chilingarian, Coarasa, Commichau, Contreras, Cortina, Costado, Curtef,
  Danielyan, Dazzi, Angelis, Delgado, {de los Reyes}, Lotto,
  Domingo-Santamar{\'i}a, Dorner, Doro, Errando, Fagiolini, Ferenc,
  Fern{\'a}ndez, Firpo, Flix, Fonseca, Font, Fuchs, Galante,
  Garc{\'i}a-L{\'o}pez, Garczarczyk, Gaug, Giller, Goebel, Hakobyan, Hayashida,
  Hengstebeck, Herrero, H{\"o}hne, Hose, Hrupec, Hsu, Jacon, Jogler, Kosyra,
  Kranich, Kritzer, Laille, Lindfors, Lombardi, Longo, L{\'o}pez, L{\'o}pez,
  Lorenz, Majumdar, Maneva, Mannheim, Mansutti, Mariotti, Mart{\'i}nez, Mazin,
  Merck, Meucci, Meyer, Miranda, Mirzoyan, Mizobuchi, Moralejo, Nieto, Nilsson,
  Ninkovic, O{\~n}a-Wilhelmi, Otte, Oya, Paneque, Panniello, Paoletti, Paredes,
  Pasanen, Pascoli, Pauss, Pegna, Persic, Peruzzo, Piccioli, Prandini,
  Puchades, Raymers, Rhode, Rib{\'o}, Rico, Rissi, Robert, R{\"u}gamer,
  Saggion, Saito, S{\'a}nchez, Sartori, Scalzotto, Scapin, Schmitt, Schweizer,
  Shayduk, Shinozaki, Shore, Sidro, Sillanp{\"a}{\"a}, Sobczynska, Stamerra,
  Stark, Takalo, Tavecchio, Temnikov, Tescaro, Teshima, Torres, Turini, Vankov,
  Vitale, Wagner, Wibig, Wittek, Zandanel, Zanin, \& Zapatero}]{mkr501_magic}
Albert, J., {et~al.} 2007, The Astrophysical Journal, 669, 862

\bibitem[{{Aleksi{\'c}} {et~al.}(2014){Aleksi{\'c}}, {Ansoldi}, {Antonelli},
  {Antoranz}, {Babic}, {Bangale}, {Barrio}, {Gonz{\'a}lez}, {Bednarek},
  {Bernardini}, {Biasuzzi}, {Biland}, {Blanch}, {Bonnefoy}, {Bonnoli},
  {Borracci}, {Bretz}, {Carmona}, {Carosi}, {Colin}, {Colombo}, {Contreras},
  {Cortina}, {Covino}, {Da Vela}, {Dazzi}, {De Angelis}, {De Caneva}, {De
  Lotto}, {Wilhelmi}, {Mendez}, {Prester}, {Dorner}, {Doro}, {Einecke},
  {Eisenacher}, {Elsaesser}, {Fonseca}, {Font}, {Frantzen}, {Fruck}, {Galindo},
  {L{\'o}pez}, {Garczarczyk}, {Terrats}, {Gaug}, {Godinovi{\'c}}, {Mu{\~n}oz},
  {Gozzini}, {Hadasch}, {Hanabata}, {Hayashida}, {Herrera}, {Hildebrand},
  {Hose}, {Hrupec}, {Idec}, {Kadenius}, {Kellermann}, {Kodani}, {Konno},
  {Krause}, {Kubo}, {Kushida}, {La Barbera}, {Lelas}, {Lewandowska},
  {Lindfors}, {Lombardi}, {Longo}, {L{\'o}pez}, {L{\'o}pez-Coto},
  {L{\'o}pez-Oramas}, {Lorenz}, {Lozano}, {Makariev}, {Mallot}, {Maneva},
  {Mankuzhiyil}, {Mannheim}, {Maraschi}, {Marcote}, {Mariotti}, {Mart{\'i}nez},
  {Mazin}, {Menzel}, {Miranda}, {Mirzoyan}, {Moralejo}, {Munar-Adrover},
  {Nakajima}, {Niedzwiecki}, {Nilsson}, {Nishijima}, {Noda}, {Orito},
  {Overkemping}, {Paiano}, {Palatiello}, {Paneque}, {Paoletti}, {Paredes},
  {Paredes-Fortuny}, {Persic}, {Poutanen}, {Moroni}, {Prandini}, {Puljak},
  {Reinthal}, {Rhode}, {Rib{\'o}}, {Rico}, {Garcia}, {R{\"u}gamer}, {Saito},
  {Saito}, {Satalecka}, {Scalzotto}, {Scapin}, {Schultz}, {Schweizer}, {Shore},
  {Sillanp{\"a}{\"a}}, {Sitarek}, {Snidaric}, {Sobczynska}, {Spanier},
  {Stamatescu}, {Stamerra}, {Steinbring}, {Storz}, {Strzys}, {Takalo},
  {Takami}, {Tavecchio}, {Temnikov}, {Terzi{\'c}}, {Tescaro}, {Teshima},
  {Thaele}, {Tibolla}, {Torres}, {Toyama}, {Treves}, {Uellenbeck}, {Vogler},
  {Zanin}, {Kadler}, {Schulz}, {Ros}, {Bach}, {Krau{\ss}}, \&
  {Wilms}}]{ic310_14}
{Aleksi{\'c}}, J., {et~al.} 2014, Science, 346, 1080

\bibitem[{{Araudo} {et~al.}(2010){Araudo}, {Bosch-Ramon}, \& {Romero}}]{abr10}
{Araudo}, A.~T., {Bosch-Ramon}, V., \& {Romero}, G.~E. 2010, A\&A, 522, A97+

\bibitem[{Araudo {et~al.}(2013)Araudo, Bosch-Ramon, \& Romero}]{abr13}
Araudo, A.~T., Bosch-Ramon, V., \& Romero, G.~E. 2013, Monthly Notices of the
  Royal Astronomical Society, 436, 3626

\bibitem[{{Barkov} {et~al.}(2012{\natexlab{a}}){Barkov}, {Aharonian},
  {Bogovalov}, {Kelner}, \& {Khangulyan}}]{babkk10}
{Barkov}, M.~V., {Aharonian}, F.~A., {Bogovalov}, S.~V., {Kelner}, S.~R., \&
  {Khangulyan}, D. 2012{\natexlab{a}}, \apj, 749, 119

\bibitem[{{Barkov} {et~al.}(2010){Barkov}, {Aharonian}, \&
  {Bosch-Ramon}}]{bab10}
{Barkov}, M.~V., {Aharonian}, F.~A., \& {Bosch-Ramon}, V. 2010, \apj, 724, 1517

\bibitem[{{Barkov} \& {Baushev}(2011)}]{bb11}
{Barkov}, M.~V., \& {Baushev}, A.~N. 2011, \na, 16, 46

\bibitem[{{Barkov} {et~al.}(2012{\natexlab{b}}){Barkov}, {Bosch-Ramon}, \&
  {Aharonian}}]{bba12}
{Barkov}, M.~V., {Bosch-Ramon}, V., \& {Aharonian}, F.~A. 2012{\natexlab{b}},
  \apj, 755, 170

\bibitem[{{Barkov} \& {Komissarov}(2016)}]{bk16}
{Barkov}, M.~V., \& {Komissarov}, S.~S. 2016, \mnras, 458, 1939

\bibitem[{{Barniol Duran} {et~al.}(2016){Barniol Duran}, {Tchekhovskoy}, \&
  {Giannios}}]{dtg16}
{Barniol Duran}, R., {Tchekhovskoy}, A., \& {Giannios}, D. 2016, ArXiv e-prints

\bibitem[{{Bednarek} \& {Protheroe}(1997)}]{bp97}
{Bednarek}, W., \& {Protheroe}, R.~J. 1997, \mnras, 287, L9

\bibitem[{{Begelman} {et~al.}(1984){Begelman}, {Blandford}, \& {Rees}}]{bbr84}
{Begelman}, M.~C., {Blandford}, R.~D., \& {Rees}, M.~J. 1984, Reviews of Modern
  Physics, 56, 255

\bibitem[{{Beskin}(2010)}]{beskin_book}
{Beskin}, V.~S. 2010, {MHD Flows in Compact Astrophysical Objects} (Springer
  Berlin Heidelberg)

\bibitem[{{Beskin} {et~al.}(1992){Beskin}, {Istomin}, \& {Parev}}]{BIP92}
{Beskin}, V.~S., {Istomin}, Y.~N., \& {Parev}, V.~I. 1992, Soviet Astronomy,
  36, 642

\bibitem[{{Beskin} \& {Nokhrina}(2006)}]{bn06}
{Beskin}, V.~S., \& {Nokhrina}, E.~E. 2006, \mnras, 367, 375

\bibitem[{{Bisnovatyi-Kogan} \& {Blinnikov}(1977)}]{BisBli77}
{Bisnovatyi-Kogan}, G.~S., \& {Blinnikov}, S.~I. 1977, \aap, 59, 111

\bibitem[{{Bisnovatyi-Kogan} \& {Lovelace}(2007)}]{BisLov07}
{Bisnovatyi-Kogan}, G.~S., \& {Lovelace}, R.~V.~E. 2007, \apjl, 667, L167

\bibitem[{{Blandford} \& {K{\"o}nigl}(1979)}]{bk79}
{Blandford}, R.~D., \& {K{\"o}nigl}, A. 1979, \apj, 232, 34

\bibitem[{{Blandford} \& {Znajek}(1977)}]{BZ77}
{Blandford}, R.~D., \& {Znajek}, R.~L. 1977, \mnras, 179, 433

\bibitem[{{Bonnoli} {et~al.}(2011){Bonnoli}, {Ghisellini}, {Foschini},
  {Tavecchio}, \& {Ghirlanda}}]{bgf11}
{Bonnoli}, G., {Ghisellini}, G., {Foschini}, L., {Tavecchio}, F., \&
  {Ghirlanda}, G. 2011, \mnras, 410, 368

\bibitem[{{Bosch-Ramon} {et~al.}(2012){Bosch-Ramon}, {Perucho}, \&
  {Barkov}}]{bpb12}
{Bosch-Ramon}, V., {Perucho}, M., \& {Barkov}, M.~V. 2012, \aap, 539, A69

\bibitem[{{Broderick} \& {Tchekhovskoy}(2015)}]{BroTch15}
{Broderick}, A.~E., \& {Tchekhovskoy}, A. 2015, \apj, 809, 97

\bibitem[{{Celotti} {et~al.}(1998){Celotti}, {Fabian}, \& {Rees}}]{cfr98}
{Celotti}, A., {Fabian}, A.~C., \& {Rees}, M.~J. 1998, \mnras, 293, 239

\bibitem[{{Cui} {et~al.}(2012){Cui}, {Yuan}, {Li}, \& {Wang}}]{cyl12}
{Cui}, Y.-D., {Yuan}, Y.-F., {Li}, Y.-R., \& {Wang}, J.-M. 2012, \apj, 746, 177

\bibitem[{{de la Cita} {et~al.}(2016){de la Cita}, {Bosch-Ramon},
  {Paredes-Fortuny}, {Khangulyan}, \& {Perucho}}]{cbp16}
{de la Cita}, V.~M., {Bosch-Ramon}, V., {Paredes-Fortuny}, X., {Khangulyan},
  D., \& {Perucho}, M. 2016, \aap, 591, A15

\bibitem[{{Falco} {et~al.}(1999){Falco}, {Kurtz}, {Geller}, {Huchra}, {Peters},
  {Berlind}, {Mink}, {Tokarz}, \& {Elwell}}]{uzc}
{Falco}, E.~E., {et~al.} 1999, \pasp, 111, 438

\bibitem[{{Ferrarese} \& {Merritt}(2000)}]{fm00}
{Ferrarese}, L., \& {Merritt}, D. 2000, \apjl, 539, L9

\bibitem[{{Gebhardt} \& {Thomas}(2009)}]{gt09}
{Gebhardt}, K., \& {Thomas}, J. 2009, \apj, 700, 1690

\bibitem[{{Gebhardt} {et~al.}(2000){Gebhardt}, {Bender}, {Bower}, {Dressler},
  {Faber}, {Filippenko}, {Green}, {Grillmair}, {Ho}, {Kormendy}, {Lauer},
  {Magorrian}, {Pinkney}, {Richstone}, \& {Tremaine}}]{gbb00}
{Gebhardt}, K., {et~al.} 2000, \apjl, 539, L13

\bibitem[{{Giannios} {et~al.}(2009){Giannios}, {Uzdensky}, \&
  {Begelman}}]{gub09}
{Giannios}, D., {Uzdensky}, D.~A., \& {Begelman}, M.~C. 2009, \mnras, 395, L29

\bibitem[{{Giannios} {et~al.}(2010){Giannios}, {Uzdensky}, \&
  {Begelman}}]{gub10}
---. 2010, \mnras, 402, 1649

\bibitem[{{Goldreich} \& {Julian}(1969)}]{gj69}
{Goldreich}, P., \& {Julian}, W.~H. 1969, \apj, 157, 869

\bibitem[{{Golkhou} {et~al.}(2015){Golkhou}, {Butler}, \&
  {Littlejohns}}]{gbl15}
{Golkhou}, V.~Z., {Butler}, N.~R., \& {Littlejohns}, O.~M. 2015, \apj, 811, 93

\bibitem[{{Gregori} {et~al.}(2000){Gregori}, {Miniati}, {Ryu}, \&
  {Jones}}]{gmr00}
{Gregori}, G., {Miniati}, F., {Ryu}, D., \& {Jones}, T.~W. 2000, \apj, 543, 775

\bibitem[{{Hayashida} {et~al.}(2015){Hayashida}, {Nalewajko}, {Madejski},
  {Sikora}, {Itoh}, {Ajello}, {Blandford}, {Buson}, {Chiang}, {Fukazawa},
  {Furniss}, {Urry}, {Hasan}, {Harrison}, {Alexander}, {Balokovi{\'c}},
  {Barret}, {Boggs}, {Christensen}, {Craig}, {Forster}, {Giommi},
  {Grefenstette}, {Hailey}, {Hornstrup}, {Kitaguchi}, {Koglin}, {Madsen},
  {Mao}, {Miyasaka}, {Mori}, {Perri}, {Pivovaroff}, {Puccetti}, {Rana},
  {Stern}, {Tagliaferri}, {Westergaard}, {Zhang}, {Zoglauer}, {Gurwell},
  {Uemura}, {Akitaya}, {Kawabata}, {Kawaguchi}, {Kanda}, {Moritani}, {Takaki},
  {Ui}, {Yoshida}, {Agarwal}, \& {Gupta}}]{hnm15}
{Hayashida}, M., {et~al.} 2015, \apj, 807, 79

\bibitem[{{Hirotani} \& {Pu}(2016)}]{hp16}
{Hirotani}, K., \& {Pu}, H.-Y. 2016, \apj, 818, 50

\bibitem[{{Hirotani} {et~al.}(2016){Hirotani}, {Pu}, {Chun-Che Lin}, {Chang},
  {Inoue}, {Kong}, {Matsushita}, \& {Tam}}]{hpc16}
{Hirotani}, K., {Pu}, H.-Y., {Chun-Che Lin}, L., {Chang}, H.-K., {Inoue}, M.,
  {Kong}, A.~K.~H., {Matsushita}, S., \& {Tam}, P.-H.~T. 2016, \apj, 833, 142

\bibitem[{{Kadler} {et~al.}(2012){Kadler}, {Eisenacher}, {Ros}, {Mannheim},
  {Els{\"a}sser}, \& {Bach}}]{ker12}
{Kadler}, M., {Eisenacher}, D., {Ros}, E., {Mannheim}, K., {Els{\"a}sser}, D.,
  \& {Bach}, U. 2012, \aap, 538, L1

\bibitem[{{Khangulyan} {et~al.}(2013){Khangulyan}, {Barkov}, {Bosch-Ramon},
  {Aharonian}, \& {Dorodnitsyn}}]{kbbad13}
{Khangulyan}, D.~V., {Barkov}, M.~V., {Bosch-Ramon}, V., {Aharonian}, F.~A., \&
  {Dorodnitsyn}, A.~V. 2013, \apj, 774, 113

\bibitem[{{Komissarov}(1994)}]{k94}
{Komissarov}, S.~S. 1994, \mnras, 266, 649

\bibitem[{{Komissarov}(2003)}]{kom03}
---. 2003, \mnras, 341, 717

\bibitem[{{Komissarov}(2004)}]{kom04}
---. 2004, \mnras, 350, 427

\bibitem[{{Komissarov} {et~al.}(2007){Komissarov}, {Barkov}, {Vlahakis}, \&
  {K{\"o}nigl}}]{kbvk07}
{Komissarov}, S.~S., {Barkov}, M.~V., {Vlahakis}, N., \& {K{\"o}nigl}, A. 2007,
  \mnras, 380, 51

\bibitem[{{Komissarov} {et~al.}(2009){Komissarov}, {Vlahakis}, {K{\"o}nigl}, \&
  {Barkov}}]{kvkb09}
{Komissarov}, S.~S., {Vlahakis}, N., {K{\"o}nigl}, A., \& {Barkov}, M.~V. 2009,
  \mnras, 394, 1182

\bibitem[{{Levinson}(2000)}]{lev00}
{Levinson}, A. 2000, Physical Review Letters, 85, 912

\bibitem[{{Levinson} {et~al.}(2005){Levinson}, {Melrose}, {Judge}, \&
  {Luo}}]{lmj05}
{Levinson}, A., {Melrose}, D., {Judge}, A., \& {Luo}, Q. 2005, \apj, 631, 456

\bibitem[{{Levinson} \& {Rieger}(2011)}]{lr11}
{Levinson}, A., \& {Rieger}, F. 2011, \apj, 730, 123

\bibitem[{{Li} {et~al.}(2009){Li}, {Yuan}, {Wang}, {Wang}, \& {Zhang}}]{lyw09}
{Li}, Y.-R., {Yuan}, Y.-F., {Wang}, J.-M., {Wang}, J.-C., \& {Zhang}, S. 2009,
  \apj, 699, 513

\bibitem[{{Lovelace}(1976)}]{lovelace76}
{Lovelace}, R.~V.~E. 1976, \nat, 262, 649

\bibitem[{{Lynden-Bell}(1969)}]{lyn69}
{Lynden-Bell}, D. 1969, \nat, 223, 690

\bibitem[{{Lyubarsky}(2005)}]{lub05}
{Lyubarsky}, Y.~E. 2005, \mnras, 358, 113

\bibitem[{{McKinney} {et~al.}(2012){McKinney}, {Tchekhovskoy}, \&
  {Blandford}}]{mtb12}
{McKinney}, J.~C., {Tchekhovskoy}, A., \& {Blandford}, R.~D. 2012, \mnras, 423,
  3083

\bibitem[{{McKinney} \& {Uzdensky}(2012)}]{mu12}
{McKinney}, J.~C., \& {Uzdensky}, D.~A. 2012, \mnras, 419, 573

\bibitem[{{Nakamura} {et~al.}(2006){Nakamura}, {McKee}, {Klein}, \&
  {Fisher}}]{nmk06}
{Nakamura}, F., {McKee}, C.~F., {Klein}, R.~I., \& {Fisher}, R.~T. 2006, \apjs,
  164, 477

\bibitem[{{Narayan} \& {Yi}(1994)}]{ny94}
{Narayan}, R., \& {Yi}, I. 1994, \apjl, 428, L13

\bibitem[{{Narayan} \& {Yi}(1995{\natexlab{a}})}]{ny95a}
---. 1995{\natexlab{a}}, \apj, 444, 231–243

\bibitem[{{Narayan} \& {Yi}(1995{\natexlab{b}})}]{ny95}
---. 1995{\natexlab{b}}, \apj, 452, 710

\bibitem[{{Narayan} \& {Yi}(1995{\natexlab{c}})}]{ny95b}
---. 1995{\natexlab{c}}, \apj, 452, 710

\bibitem[{{Neronov} \& {Aharonian}(2007)}]{NerAha07}
{Neronov}, A., \& {Aharonian}, F.~A. 2007, \apj, 671, 85

\bibitem[{{Petropoulou} {et~al.}(2016){Petropoulou}, {Giannios}, \&
  {Sironi}}]{pgs16}
{Petropoulou}, M., {Giannios}, D., \& {Sironi}, L. 2016, \mnras, 462, 3325

\bibitem[{{Pittard} {et~al.}(2010){Pittard}, {Hartquist}, \& {Falle}}]{phf10}
{Pittard}, J.~M., {Hartquist}, T.~W., \& {Falle}, S.~A.~E.~G. 2010, \mnras,
  405, 821

\bibitem[{{Porth} \& {Komissarov}(2015)}]{pk15}
{Porth}, O., \& {Komissarov}, S.~S. 2015, \mnras, 452, 1089

\bibitem[{{Pozanenko} \& {Loznikov}(2002)}]{pl02}
{Pozanenko}, A., \& {Loznikov}, V. 2002, in {Lighthouses of the Universe: The
  Most Luminous Celestial Objects and Their Use for Cosmology}, ed.
  M.~{Gilfanov}, R.~{Sunyeav}, \& E.~{Churazov}, 194

\bibitem[{{Rieger}(2011)}]{rieger11}
{Rieger}, F.~M. 2011, International Journal of Modern Physics D, 20, 1547

\bibitem[{{Rieger} \& {Aharonian}(2008)}]{RieAha08}
{Rieger}, F.~M., \& {Aharonian}, F.~A. 2008, \aap, 479, L5

\bibitem[{{Rieger} \& {Volpe}(2010)}]{rv10}
{Rieger}, F.~M., \& {Volpe}, F. 2010, \aap, 520, A23

\bibitem[{{Romoli} {et~al.}(2017){Romoli}, {Taylor}, \& {Aharonian}}]{rta17}
{Romoli}, C., {Taylor}, A.~M., \& {Aharonian}, F. 2017, Astroparticle Physics,
  88, 38

\bibitem[{{Ruderman} \& {Sutherland}(1975)}]{RudSut75}
{Ruderman}, M.~A., \& {Sutherland}, P.~G. 1975, \apj, 196, 51

\bibitem[{Ruffini \& Wilson(1975)}]{ruffini75}
Ruffini, R., \& Wilson, J.~R. 1975, Phys. Rev. D, 12, 2959

\bibitem[{{Salpeter}(1964)}]{sal64}
{Salpeter}, E.~E. 1964, \apj, 140, 796

\bibitem[{{Shakura} \& {Sunyaev}(1973)}]{ss73}
{Shakura}, N.~I., \& {Sunyaev}, R.~A. 1973, \aap, 24, 337

\bibitem[{{Sijbring} \& {de Bruyn}(1998)}]{SijBru98}
{Sijbring}, D., \& {de Bruyn}, A.~G. 1998, \aap, 331, 901

\bibitem[{{Sironi} {et~al.}(2016){Sironi}, {Giannios}, \&
  {Petropoulou}}]{sgp16}
{Sironi}, L., {Giannios}, D., \& {Petropoulou}, M. 2016, \mnras, 462, 48

\bibitem[{{Stern} \& {Poutanen}(2006)}]{sp06}
{Stern}, B.~E., \& {Poutanen}, J. 2006, \mnras, 372, 1217

\bibitem[{{Striani} {et~al.}(2010){Striani}, {Vercellone}, {Tavani},
  {Vittorini}, {D'Ammando}, {Donnarumma}, {Pacciani}, {Pucella}, {Bulgarelli},
  {Trifoglio}, {Gianotti}, {Giommi}, {Argan}, {Barbiellini}, {Caraveo},
  {Cattaneo}, {Chen}, {Costa}, {De Paris}, {Del Monte}, {Di Cocco}, \&
  {Evangelista}}]{agile10}
{Striani}, E., {et~al.} 2010, \apj, 718, 455

\bibitem[{{Sturrock}(1971)}]{stu71}
{Sturrock}, P.~A. 1971, \apj, 164, 529

\bibitem[{{Timokhin}(2010)}]{timokhin10}
{Timokhin}, A.~N. 2010, \mnras, 408, 2092

\bibitem[{{Timokhin} \& {Arons}(2013)}]{ta13}
{Timokhin}, A.~N., \& {Arons}, J. 2013, \mnras, 429, 20

\bibitem[{{Urry} \& {Padovani}(1995)}]{up95}
{Urry}, C.~M., \& {Padovani}, P. 1995, \pasp, 107, 803

\bibitem[{{Vovk} \& {Babi{\'c}}(2015)}]{vb15}
{Vovk}, I., \& {Babi{\'c}}, A. 2015, \aap, 578, A92

\bibitem[{{Wang} \& {Zhou}(2009)}]{wz09}
{Wang}, C.-C., \& {Zhou}, H.-Y. 2009, \mnras, 395, 301

\bibitem[{{Zel'dovich} \& {Novikov}(1966)}]{zn66}
{Zel'dovich}, Y.~B., \& {Novikov}, I.~D. 1966, Soviet Physics Uspekhi, 8, 522

\end{thebibliography}

\end{document}